\documentclass{iopart}

\usepackage{graphicx}
\usepackage{amssymb}
\usepackage{psfrag}

\begin{document}

\newcommand \be {\begin{equation}}
\newcommand \ee {\end{equation}}
\newcommand \bea {\begin{eqnarray}}
\newcommand \eea {\end{eqnarray}}

\title[Microscopic derivation of hydrodynamic equations for self-propelled
particles]{Hydrodynamic equations for self-propelled particles:
microscopic derivation and stability analysis}

\author{Eric Bertin,$^{1,2}$ Michel Droz$^2$ and Guillaume Gr\'egoire$^3$}

\address{$^1$ Universit\'e de Lyon, Laboratoire de Physique, ENS Lyon, CNRS,
46 All\'ee d'Italie, F-69007 Lyon\\
$^2$Department of Theoretical Physics,
University of Geneva, CH-1211 Geneva 4, Switzerland\\
$^3$Laboratoire Mati\`ere et Syst\`emes Complexes (MSC), UMR 7057,
CNRS-Universit\'e 
Paris-Diderot, F-75205 Paris Cedex 13, France
}

\begin{abstract}
Considering a gas of self-propelled particles with binary interactions,
we derive the hydrodynamic equations governing the density and velocity
fields from the microscopic dynamics, in the framework of the associated
Boltzmann equation. Explicit expressions for the transport coefficients
are given, as a function of the microscopic parameters of the model.
We show that the homogeneous state with zero hydrodynamic velocity is
unstable above a critical density
(which depends on the microscopic parameters),
signaling the onset of a collective motion.
Comparison with numerical simulations on a standard model of self-propelled
particles shows that the phase diagram we obtain is robust, in the sense that
it depends only slightly on the precise definition of the model.
While the homogeneous flow is found to be stable far from the transition line,
it becomes unstable with respect to finite-wavelength perturbations
close to the transition, implying a non trivial spatio-temporal structure
for the resulting flow.
We find solitary wave solutions of the hydrodynamic
equations, quite similar to the stripes reported in direct
numerical simulations of self-propelled particles.
\end{abstract}
\pacs{05.70.Ln, 05.20.Dd, 64.60.Cn}

%%%%%%%%%%%%%%%%%%%%%%%%%%%%%%%%%%%%%%%%%%%%%%%%%%%%%%%%%%%%%%%%%%%%

\section{Introduction}

In the recent years, a lot of effort has been expended with the aim of
explaining the collective behaviour of living systems \cite{Toner_rev}.
Such collective behaviours can be observed on many different scales
including mammal herds~\cite{Parrish1997}, crowds of
pedestrians~\cite{Helbing_nat,Helbing}, bird flocks~\cite{Feare1984},
fish schools~\cite{Hubbard2004}, insect swarms~\cite{Rauch1995},
colonies of bacteria~\cite{BenJacob1995}, molecular
motors~\cite{Harada1987,Badoual2002} and even interacting
robots~\cite{Sugarawa2002}.  It turns out that the collective
properties of such systems seem to be quite robust and
universal. Accordingly, this field attracted the interest of the
statistical physics community with the challenge of introducing
minimal models that could capture the emergence of collective
behaviour.  One important class of models consists of the so-called
self-propelled particles models, for which the onset of collective
motion without a leader is present.  Vicsek \emph{et
  al.}~\cite{Vicsek1995,Czirok1997} introduced a simple model defined
on a continuous plane, where agents (or animals) are represented as
point particles with a velocity of constant amplitude. Noisy
interaction rules tend to align the velocity of any given particle
with its neighbors. Extensive numerical simulations of this model have
been performed~\cite{Gregoire2004,Chate2008}, showing the presence of
a phase transition from a disordered state, at high enough noise, to a
state with collective motion.  A different approach is to consider the
problem at a coarsed-grained level and to describe the dynamics in
terms of hydrodynamic fields. The equations governing the evolution of
these hydrodynamic fields can be either postulated phenomenologically
\cite{Csahok2002}, on the basis of symmetry and conservation laws
considerations \cite{Toner1995,Toner1998}, or derived from specific
microscopic models \cite{Bertin2006,Baskaran2008}. The equations of
motion of the hydrodynamic field are derived from the microscopic
model through a Boltzman approach.

Following an earlier publication \cite{Bertin2006},
the motivation of the present work is to derive, from a microscopic model,
the hydrodynamic equations describing at a coarse-grained level the flow
of self-propelled particles (SPP), and to compare the resulting description
with numerical simulations of an agent-based model of SPP.
The analytical framework we use is that of the Boltzmann equation.
Accordingly, a suitable microscopic model for such a treatment is a
continuous time model with interactions reducing to binary collisions.
In order to show that the most salient features of the coarse-grained
analytical description are not specific to a binary collision model,
we use for the numerical simulations a standard agent-based model
\cite{Vicsek1995,Czirok1997},
that has been well characterized in the litterature 
\cite{Gregoire2004,Chate2008}.
Note that some comparison with numerical simulations of an agent-based model
with binary interaction have already been presented in Ref.~\cite{Bertin2006}.

%%%%%%%%%%%%%%%%%%%%%%%%%%%%%%%%%%%%%%%%%%%%%%%%%%%%%%%%%%%%%%%%%%%%%

\section{Microscopic models of interacting self-propelled particles}

%%%%%%%%%%%
\subsection{Definition of the models}\label{sec-definition}

\subsubsection{Continuous time model with binary collisions.}

Following Ref.~\cite{Bertin2006},
we introduce a simple model that captures the essential physics
of assemblies of self-propelled particles, while being suitable
for a description in terms of a Boltzmann equation.
We consider the evolution of self-propelled point-like particles on a
two-dimensional plane. The displacement of each particle $i$ is governed
by a velocity vector $\mathbf{v}_i$. In order to account for the
self-propelling property, we assume that the modulus of the velocity
vector is fixed to a value $v_0$, identical for all the particles,
so that only the direction of the vector plays a role in the dynamics.
The relevant dynamical variables are then the angles $\theta_i$ that
the vectors $\mathbf{v}_i$ form with a fixed reference direction.
It is important to note at this stage, that fixing the modulus of the
velocity breaks the Galilean invariance of the system. Hence one should
not expect that the eventually obtained hydrodynamic equations obey
such an invariance, contrary to what happens in usual flows.

Apart from the ballistic evolution according to their velocity vector,
particles also experience stochastic events that punctuate their dynamics.
These stochastic events are of two different types. The simplest ones are
self-diffusion events, that is, the angle $\theta$ of an isolated particle
changes, with a probability $\lambda$ per unit time, to
$\theta'=\theta+\eta$, where $\eta$ is a noise with distribution $p_0(\eta)$
and variance $\sigma_0^2$. In the following, we consider a Gaussian
distribution for $p_0(\eta)$, also taking into account the periodicity
of $\theta$.
This type of stochastic events lead to a diffusive behaviour at large scale,
thus preventing the system from having a trivial (pseudo)collective motion,
of purely ballistic nature.
To drive the system into an organised state where a genuine collective
motion sets in, one has to introduce interactions between the particles.
Given that we wish to use a Boltzmann approach to study the model,
it is natural to consider binary interactions between particles.
These binary interactions are introduced as follows. When two particles
get closer than a threshold distance $d_0$, their velocity angle
$\theta_1$ and $\theta_2$ are changed into $\theta_1'$ and $\theta_2'$
according to:
\be
\theta_1'=\overline{\theta}+\eta_1 [2\pi], \qquad
\theta_2'=\overline{\theta}+\eta_2 [2\pi],
\ee
where $\overline{\theta}$ is defined by:
\be
\overline{\theta} = \arg\left( e^{i\theta_1} + e^{i\theta_2} \right).
\ee
The noises $\eta_1$ and $\eta_2$ are independent Gaussian variables with
variances $\sigma^2$. Note that $\sigma^2$ may differ from the variance
$\sigma_0^2$ of the noise associated to the self-diffusion of particles.

\subsubsection{Agent-based model for numerical simulations.}

In order to compare the results of the analytical approach based on the
binary collision model to direct numerical simulations, we use a slight
generalization of the standard Vicsek model \cite{Vicsek1995,Czirok1997}.
The motivation for simulating numerically a model different from
the one we used in the analytical approach is twofold.
First, the Vicsek model has been thoroughly characterized in the
literature \cite{Czirok1997,Gregoire2004,Chate2008},
making it a useful benchmark for comparison.
Second, and most importantly, a model with continuous time dynamics and
binary collisions is well-suited for a Boltzmann equation approach,
but very inefficient from the point of view of direct numerical simulations.
In constrast, the Vicsek model, with a discrete time
dynamics and multi-neighbour interactions, is much more efficient to simulate.

The agent-based model we consider consists in $N$ particles
on a two-dimensional space of area $L\times L$, with periodic boundary
conditions.
Each particle $j$ at any instant $t$ has a constant modulus speed
$v_0$. This property allows the mapping of velocity on complex
numbers. Then a particle is located by a two-coordinate vector
$\mathbf{x}_j ^t$ and an angle $\vartheta_j ^t$ which gives its speed
direction. We define the vicinity $\mathcal{V}_j^t$ of $j$ at time $t$
as the disk centred on $j$ with a radius $d_0$. Then the direction of
$j$ at the next instant $t+\Delta t$ is simply the direction of the
averaged speed over all particles which are embedded in its vicinity,
including $j$ itself, up to a noise term. If there is no neighbour in
the disk of interaction, self-diffusion occurs randomly:
\bea
\theta_j^{t+\Delta t}&=& \left\lbrace
\begin{array}{l}
\arg\left[\sum_{k\in\mathcal{V}_j^t}e^{i\theta_k^t}\right]+\eta\xi_j^t ,\; 
\mathrm{if}\; \mathcal{V}_j^t\ne\{j\},\\ \\
\theta_j^t+\eta_0\xi_j^t,\;\mathrm{with\; probability}\;\lambda\Delta t,\;
\mathrm{if}\; \mathcal{V}_j^t=\{j\} ,\\ \\
\theta_j^t,\;\mathrm{with\; probability}\; 1-\lambda\Delta t,\;
\mathrm{if}\; \mathcal{V}_j^t=\{j\} ,\\
\end{array}
\right.\\
\mathbf{x}_i^{t+\Delta t}&=&\mathbf{x}_i^t+v_0
\mathbf{e}(\theta_j^{t+\Delta t})\Delta t,
\eea
where $\mathbf{e}(\theta)$ is the unit vector of direction $\theta$.
The parameters $\eta$ and $\eta_0$ are the noise amplitudes for collision
and self-diffusion respectively.
The random number $\xi_j^t$ is uncorrelated in time and
from one particle to another. Its distribution is flat on
$[-\pi,\pi]$.
The slight generalization with respect to the standard Vicsek model
consists in the introduction of the parameter $\lambda$, which
characterizes the probability of self-diffusion per unit time.
In the original model, $\lambda \Delta t=1$.
Note that, whenever possible, we have defined
the agent-based model with notations consistent with that of the
binary collision model, in order to facilitate comparison between the
two models.

The Vicsek model has been studied in details in the literature
\cite{Vicsek1995,Gregoire2004,Chate2008}.
A transition toward collective motion
has been reported in early studies \cite{Vicsek1995}, and later shown
to exhibit strong finite size effects \cite{Gregoire2004}.
In \ref{app-num-model}, we recall the methodology used to study
the transition, and in particular the finite size scaling effects.

%%%%%%%%%%%
\subsection{Dimensionless parameters}

A first step in the understanding of the models is to identify the
relevant dimensionless parameters and the possible regimes.  Let us
first consider the different length scales appearing in this problem: the
interaction range $d_0$, the ballistic length
$\ell_\mathrm{bal}=v_0/\lambda$, and the typical distance between particles
$\ell_\mathrm{pp}=1/\sqrt{\rho}$.  With these three different lengths,
one can form the following dimensionless numbers $H$ and $B$:
\be
H = \frac{\ell_\mathrm{pp}}{d_0} = \frac{1}{d_0 \sqrt{\rho}} \;, \qquad
B = \frac{\ell_\mathrm{bal}}{d_0} = \frac{v_0}{d_0 \lambda} \;.
\ee
$H$ characterises whether a system is diluted ($H\gg 1$) or dense. One
can see $B$ as the relative weight of stand-alone flight over
interaction. If $B$ is large, ballistic flight is more important than
collision and we can expect that particles are less correlated
locally.

These numbers turn out to play an important role in the identification
of the regimes of validity of the approximations we use, as seen in
the following. The model also exhibits different behaviours for the
different regimes which are defined by these numbers.  At fixed noise
intensity and fixed $B$, a more (resp. less) dense system is expected
to move (resp. not) in a collective manner. At fixed noise and fixed
dilution $H$, increasing $B$ makes the flights more ballistic,
which should favor collective motion. So one can guess that a relevant
control parameter will be a combination of $H$ and $B$
(see section~\ref{sec-statsol}).

%%%%%%%%%%%
\subsection{Summary of the main results}

The paper is organised as follows. Section~\ref{sec-eq-hydro} is devoted
to the derivation from the Boltzmann equation, through a specific
approximation scheme, of the hydrodynamic equations for the continuous time
binary collision model.
Section~\ref{sec-phase-diag} deals with the analysis of the phase
diagram of the binary collision model, by looking at the
stationary homogeneous solutions and studying their linear stability.
A transition toward collective motion is observed, but the spatially
homogeneous motion turns out to be unstable in the validity
domain of the hydrodynamic equations, namely close to the transition line.
A comparison with the agent-based model is presented, showing
that the transition lines of both models are qualitatively similar,
and share some quantitative properties.
Then, Section~\ref{sec-beyond-valid} investigates the behaviour of
the binary collision model beyond the strict domain of validity of
the hydrodynamic equations. A direct stability analysis shows that
far from the transition line, the spatially homogeneous motion is stable.
We then test whether the hydrodynamic equations could be used, in this
domain, as a semi-quantitative description. We find that the
restabilization phenomenon is indeed observed in the hydrodynamic equations,
although the predicted location of the transition line between stable
and unstable motion does not match quantitatively with a perturbative
treatment of the kinetic theory.
We also show that there exist solitary wave solutions of the hydrodynamic
equations, that resemble the travelling stripes of higher density observed
in the agent-based model.
Finally, Section~\ref{sec-discussion} discusses the main conclusions
and perspectives of the present work. Some technical aspects related
to the agent-based model and to the stability analysis of the homogeneous
motion are reported in \ref{app-num-model} and
\ref{app-stability} respectively.

%%%%%%%%%%%%%%%%%%%%%%%%%%%%%%%%%%%%%%%%%%%%%%%%%%%%%%%%%%%%%%%%%%%%%%%%%

\section{Boltzmann approach and hydrodynamic equations}
\label{sec-eq-hydro}

%%%%%%%%%%%
\subsection{Description in terms of Boltzmann equation}

One of the main goals of this work is to derive analytically from the
microscopic dynamics, within an appropriate approximation scheme, the equations
governing the evolution of the hydrodynamic fields, namely the density
and velocity fields. A standard approach to obtain these hydrodynamic
equations is to write, as a first step, the Boltzmann equation
describing the evolution of the one-particle probability distribution
in phase-space (i.e., the probability that a particle is at a given point,
with a given velocity), and then to derive hydrodynamic equations
by computing the first moments of the Boltzmann equation. Note however
that such a procedure often yields a hierarchy of equations, so that a
closure assumption has to be used.

Let us start by deriving the Boltzmann equation for the above model.
This equation relies on the standard assumption that the gas is diluted,
meaning that the typical distance $\ell_\mathrm{pp}$ between particles
is large compared to the interaction distance $d_0$, that is $H \gg 1$.
In the present context, one also needs to assume that the ballistic
distance $\ell_\mathrm{bal}$ is much larger than $d_0$, namely $B \gg
1$. It ensures that there is no memory effect from one collision
to the other.
The Boltzmann equation governs the evolution of the distribution
$f(\mathbf{r},\theta,t)$, that gives the probability that a particle
is at point $\mathbf{r}$ with a velocity along the direction defined by the
angle $\theta$. On general grounds, this equation can be written as
\be \label{eq-boltz}
\frac{\partial f}{\partial t}(\mathbf{r},\theta,t) + v_0\, \mathbf{e}(\theta)
\cdot \nabla f(\mathbf{r},\theta,t) = I_{\mathrm{dif}}[f] +
I_{\mathrm{col}}[f,f].
\ee
The different terms in the equation can be interpreted as follows.
The second term in the l.h.s.~corresponds to the ballistic motion of particles
between two stochastic events (self-diffusion or collision).
In the r.h.s., the term $I_{\mathrm{dif}}[f]$ accounts for the self-diffusion
events, and it reads
\bea \label{Idif}
I_{\mathrm{dif}}[f] &=& -\lambda f(\mathbf{r},\theta,t)
+ \lambda \int_{-\pi}^{\pi} d\theta' \int_{-\infty}^{\infty} d\eta\,
p_0(\eta)\\
&& \quad \times
\sum_{m=-\infty}^{\infty} \delta(\theta'+\eta-\theta+2m\pi)
f(\mathbf{r},\theta',t).
\nonumber
\eea
The sum of $\delta$-distributions accounts for the periodicity of angles.
Finally, the term $I_{\mathrm{col}}[f,f]$ describes the effects of collisions.
It can be derived in the following way. A collision between two particles
occurs if their relative distance becomes less than $d_0$. Although the
two particles a priori play a symmetric role, it is convenient to choose
one particle, and to observe the situation in the referential of this
particle --say particle $1$. In this frame, the velocity of particle
$2$ is $\tilde{\mathbf{v}}_2=v_0 (\mathbf{e}(\theta_2)-\mathbf{e}(\theta_1))$.
Hence, in order to collide with particle $1$ between $t$ and $t+dt$,
particle $2$ has to lie at time $t$ (in the referential of
particle $1$) in a rectangle of length
$|\tilde{\mathbf{v}}_2|dt$ and of width $2d_0$.
Coming back to the laboratory frame, this rectangle deforms
into a parallelogram, but keeps the same surface, given by
$2d_0 v_0 |\mathbf{e}(\theta_2)-\mathbf{e}(\theta_1)| dt$.
The collision term $I_{\mathrm{col}}[f,f]$ is then obtained
from the bilinear functional $I_{\mathrm{col}}[g,h]$:
\bea \label{Icol}
I_{\mathrm{col}}[g,h] &=&  -2d_0 v_0 g(\mathbf{r},\theta,t)
\int_{-\pi}^{\pi} d\theta'\,
|\mathbf{e}(\theta')-\mathbf{e}(\theta)| h(\mathbf{r},\theta',t)\\
\nonumber
&+& 2d_0 v_0 \int_{-\pi}^{\pi} d\theta_1 \int_{-\pi}^{\pi} d\theta_2
\int_{-\infty}^{\infty} d\eta\, p(\eta)\,
|\mathbf{e}(\theta_2)-\mathbf{e}(\theta_1)|\\
\nonumber
&\times& g(\mathbf{r},\theta_1,t) h(\mathbf{r},\theta_2,t)
\sum_{m=-\infty}^{\infty} \delta(\overline{\theta}+\eta-\theta+2m\pi),
\eea
with again the notation $\overline{\theta}=\arg(e^{i\theta_1}
+e^{i\theta_2})$, and where $g$ and $h$ are arbitrary phase-space
distributions.

It is straightforward to check that the uniform one-particle distribution
$f_0(\mathbf{r},\theta,t)=\rho_0/2\pi$, associated to a uniform density
of particles $\rho_0$, is a stationary solution of the Boltzmann equation,
for any values of the noise parameters $\sigma$ and $\sigma_0$, since each
term in Eq.~(\ref{eq-boltz}) vanishes independently.
If a transition to a state with collective motion occurs, another
distribution should be a steady-state solution of the Boltzmann equation.
Yet, finding this non-trivial distribution through non-perturbative
analytical method is a hard task. One could turn to numerical approaches,
but we would rather like to obtain analytical results, at least in some
specific regime.
We thus use in the following an alternative approach, which consists
in deriving hydrodynamic equations for the density and velocity
fields from the Boltzmann equation, in the limit of small hydrodynamic
velocity. A stability analysis can then be performed on these hydrodynamic
equations in order to check the onset of collective motion.

%%%%%%%%%%%
\subsection{Derivation of the hydrodynamic equations}

\subsubsection{Hydrodynamic fields and continuity equation.}

The hydrodynamic fields are on the one hand the density field:
\be
\rho(\mathbf{r},t) = \int_{-\pi}^{\pi} d\theta\, f(\mathbf{r},\theta,t),
\ee
and on the other hand the velocity field:
\be
\mathbf{u}(\mathbf{r},t) = \frac{v_0}{\rho(\mathbf{r},t)}
\int_{-\pi}^{\pi} d\theta\, f(\mathbf{r},\theta,t)\, \mathbf{e}(\theta).
\ee
The equations governing the evolution of these hydrodynamic fields are
derived by taking the successive moments of the Boltzmann equation.
A simple integration of Eq.~(\ref{eq-boltz}) over $\theta$ directly
leads the evolution equation for $\rho(\mathbf{r},t)$:
\be \label{continuity}
\frac{\partial \rho}{\partial t} + \nabla \cdot (\rho \mathbf{u}) = 0,
\ee
which is nothing but the usual continuity equation accounting for the
conservation of the number of particles.

\subsubsection{Angular Fourier expansion of the phase-space distribution.}

The derivation of the evolution equation for the velocity field is actually
much more complicated, and one has to resort to approximation schemes.
As $f(\mathbf{r},\theta,t)$ is a periodic function of $\theta$,
it is convenient to work with its Fourier series expansion, defined as:
\be
\hat{f}_k(\mathbf{r},t) = \int_{-\pi}^{\pi} d\theta \,
f(\mathbf{r},\theta,t)\, e^{ik\theta}.
\ee
Conversely, $f(\mathbf{r},\theta,t)$ can be expressed as a function of
the Fourier coefficients through the relation:
\be \label{Fourier-exp}
f(\mathbf{r},\theta,t) = \frac{1}{2\pi} \sum_{k=-\infty}^{\infty}
\hat{f}_k(\mathbf{r},t) \, e^{-ik\theta}.
\ee
In this framework, the uniform distribution
$f_0(\mathbf{r},\theta,t)=(2\pi)^{-1}\rho_0$
corresponds to $\hat{f}_k(\mathbf{r},t)= (2\pi)^{-1}\rho_0 \,\delta_{k,0}$.

Let us use as a basis of the plane
the two orthogonal vectors $\mathbf{e}_1=\mathbf{e}(0)$
and $\mathbf{e}_2=\mathbf{e}(\pi/2)$. The components of $\mathbf{e}(\theta)$
in this basis are obviously $e_1(\theta)=\cos \theta$ and
$e_2(\theta)=\sin \theta$.
In order to obtain an evolution equation for the velocity field, we
multiply Eq.~(\ref{eq-boltz}) by $\mathbf{e}(\theta)$ and integrate
over $\theta$; one gets in tensorial notations ($j=1$, $2$):
\bea \label{integ-boltz}
&&\!\!\!\!\!\!\frac{\partial}{\partial t} \int_{-\pi}^{\pi} d\theta\,
e_j(\theta) f(\mathbf{r},\theta,t) + v_0 \sum_{l=1}^2
\frac{\partial}{\partial x_l}\int_{-\pi}^{\pi}
d\theta\, e_j(\theta) e_l(\theta) f(\mathbf{r},\theta,t) = \\ \nonumber
&& \qquad \qquad \qquad \int_{-\pi}^{\pi} d\theta\, e_j(\theta)
\left( I_{\mathrm{dif}}[f] + I_{\mathrm{col}}[f,f] \right).
\eea
To proceed further, it is convenient
to identify complex numbers with two-dimensional vectors, in such a way
that $\mathbf{e}(\theta)$ is mapped onto $e^{i\theta}$.
Then, in the same way, $v_0 \hat{f}_1(\mathbf{r},t)$
is associated to the momentum field
$\mathbf{w}(\mathbf{r},t) = \rho(\mathbf{r},t)\, \mathbf{u}(\mathbf{r},t)$.
Hence, we wish to rewrite Eq.~(\ref{integ-boltz}) in such complex
notations. For later use, we shall write it in a slightly more general form,
replacing $e^{i\theta}$ with $e^{ik\theta}$ ($k$ being an integer):
\bea \label{complex-boltz}
&&\frac{\partial}{\partial t} \int_{-\pi}^{\pi} d\theta\, e^{ik\theta}
f(\mathbf{r},\theta,t) + v_0 \sum_{\ell=1}^2 \frac{\partial}{\partial x_l}
\int_{-\pi}^{\pi} d\theta
\, e^{ik\theta} e_l(\theta) f(\mathbf{r},\theta,t) = \\ \nonumber
&& \qquad \qquad \qquad
\int_{-\pi}^{\pi} d\theta\, e^{ik\theta} \left( I_{\mathrm{dif}}[f] +
I_{\mathrm{col}}[f,f] \right).
\eea 
Eq.~(\ref{integ-boltz}) is recovered for $k=1$, up to the mapping between
complex number and two-dimensional vectors.
The first term in the l.h.s.~is simply $\partial \hat{f}_k/\partial t$.
The r.h.s.~of Eq.~(\ref{complex-boltz}) is computed by inserting the
Fourier series expansion (\ref{Fourier-exp}) into Eqs.~(\ref{Idif})
and (\ref{Icol}). After a rather straightforward calculation, one finds:
\bea \label{rhs-complex}
&&\int_{-\pi}^{\pi} d\theta\,  e^{ik\theta} \left( I_{\mathrm{dif}}[f] +
I_{\mathrm{col}}[f] \right) =
- \lambda \left(1-e^{-k^2 \sigma_0^2/2}\right) \hat{f}_k(\mathbf{r},t)
\\ \nonumber
&& \qquad \qquad
- \frac{2}{\pi}d_0 v_0 \sum_{q=-\infty}^{\infty}
\left( I_q - e^{-k^2 \sigma^2/2} I_{q-k/2} \right)
\hat{f}_q(\mathbf{r},t) \hat{f}_{k-q}(\mathbf{r},t),
\eea
where the coefficients $I_q$ are defined as:
\be
I_q = \int_{-\pi}^{\pi} d\theta \, \left| \sin \frac{\theta}{2}\right|
\, \cos q\theta.
\ee
From this definition, it is obvious that $I_{-q}=I_q$.
For integer $q$, $I_q$ is given by:
\be
I_q = \frac{4}{1-4q^2},
\ee
while for half-integer $q=m+\frac{1}{2}$ ($m$ integer) one has:
\bea
I_{\frac{1}{2}} = I_{-\frac{1}{2}} = 2, \\
I_{m+\frac{1}{2}} = \frac{1}{m(m+1)} \left[ (-1)^m (2m+1)-1 \right],
\qquad m \ne -1, 0.
\eea
The second term in the l.h.s.~of Eq.~(\ref{complex-boltz})
can be evaluated as follows.
For $l=1,2$ and $k$ integer, let us define the complex quantity
$Q_l^{(k)}(\mathbf{r},t)$ as:
\be
Q_l^{(k)}(\mathbf{r},t) = \int_{-\pi}^{\pi} d\theta
\, e^{ik\theta} e_l(\theta) f(\mathbf{r},\theta,t).
\ee
The following relations are then easily obtained:
\bea \label{eq-Qk}
Q_1^{(k)}(\mathbf{r},t) &=&
\frac{1}{2} [\hat{f}_{k+1}(\mathbf{r},t)+\hat{f}_{k-1}(\mathbf{r},t)],\\
Q_2^{(k)}(\mathbf{r},t) &=&
\frac{1}{2i} [\hat{f}_{k+1}(\mathbf{r},t)-\hat{f}_{k-1}(\mathbf{r},t)].
\eea

\subsubsection{Velocity field equation in the small velocity regime.}

Up to now, the calculations made are exact, apart from the
approximations underlying the Boltzmann equation. 
As already mentioned, the Fourier coefficient $\hat{f}_0(\mathbf{r},t)$
is nothing but the density field $\rho(\mathbf{r},t)$, and
$\hat{f}_1(\mathbf{r},t)$ can be mapped onto the momentum field
$\mathbf{w}(\mathbf{r},t)$ through
the identification of complex numbers with two-dimensional vectors.
A similar mapping also holds for $\hat{f}_{-1}(\mathbf{r},t)$, which is
the complex conjugate of $\hat{f}_1(\mathbf{r},t)$. In contrast,
Fourier coefficient $\hat{f}_k(\mathbf{r},t)$ with $|k|>1$ cannot be
mapped onto the hydrodynamic fields. As it turns out
that such coefficients appear both
in the expression of $Q_l^{(k)}(\mathbf{r},t)$
and in the r.h.s.~of Eq.~(\ref{complex-boltz}),
an approximation scheme has to be found in order to obtain from
Eq.~(\ref{complex-boltz}) a closed hydrodynamic equation, involving only
the fields $\rho(\mathbf{r},t)$ and $\mathbf{u}(\mathbf{r},t)$.

In the following, we assume that the distribution $f(\mathbf{r},\theta,t)$
is close to an isotropic distribution, namely, it
depends only slightly on $\theta$.
This amounts to assuming that the hydrodynamic velocity
is much smaller than the velocity of individual particles.
In terms of Fourier coefficients, the hydrodynamic velocity
is given by
$||\mathbf{u}(\mathbf{r},t)||=v_0|\hat{f}_1(\mathbf{r},t)|/\rho(\mathbf{r},t)$.
We introduce a small parameter $\epsilon$ such that
$||\mathbf{u}(\mathbf{r},t)||=\mathcal{O}(\epsilon)$.
For instance, $\epsilon$ can be chosen as $\overline{u}/v_0$,
where $\overline{u}$ is the spatial average of $||\mathbf{u}(\mathbf{r},t)||$
at some initial time $t=t_0$.
Then the key assumption we use to build an approximation scheme is
\be \label{eq-scal-ansatz}
\hat{f}_k(\mathbf{r},t) = \mathcal{O}(\epsilon^{|k|}).
\ee
Such a scaling ansatz is consistent with the property
$\hat{f}_{-k}(\mathbf{r},t) = \hat{f}_k(\mathbf{r},t)^*$, with the
scaling properties of $\hat{f}_0(\mathbf{r},t)$ and $\hat{f}_1(\mathbf{r},t)$,
and with Eq.~(\ref{rhs-complex}).
We shall identify more precisely in Section~\ref{sec-valid} 
the validity domain of this scaling ansatz, and thus of the hydrodynamic
equations we will derive from it.

Using the above scaling ansatz, the sum in the r.h.s.~of
Eq.~(\ref{rhs-complex}), for $k=1$, can be truncated, only keeping
terms with $q=0$, $1$ or $2$, that are at most of order $\epsilon^3$, while
discarding the other terms, being of higher order in $\epsilon$.
Gathering all terms,
one obtains the following equation for the evolution of $\hat{f}_1$
(we drop the explicit dependence upon $\mathbf{r}$ and $t$ to simplify
the notations):
\bea \nonumber
&& \frac{\partial \hat{f}_1}{\partial t}
+ \frac{v_0}{2} \frac{\partial}{\partial x_1}(\hat{f}_2+\rho)
+ \frac{v_0}{2i} \frac{\partial}{\partial x_2}(\hat{f}_2-\rho)
= \\ \nonumber
&& \qquad \qquad
-\left[\lambda \left(1-e^{-\sigma_0^2/2}\right)+\frac{8}{\pi}d_0 v_0
\left(\frac{2}{3}-e^{-\sigma^2/2} \right)\rho \right] \hat{f}_1 \\
&& \qquad \qquad
-\frac{8}{\pi}d_0 v_0 \left(e^{-\sigma^2/2}-\frac{2}{5}\right)
\hat{f}_1^* \hat{f}_2.
\eea
Hence, the resulting equation involves $\hat{f}_0=\rho$, $\hat{f}_1$ and
$\hat{f}_2$.
Accordingly, it turns out that one needs to find a closure relation
to express $\hat{f}_2$ as a function of $\hat{f}_0$ and $\hat{f}_1$
(or, equivalently, in terms of $\rho$ and $\mathbf{u}$).
Such a relation is given by the evolution equation for $\hat{f}_2$,
that is, Eq.~(\ref{complex-boltz}) with $k=2$. From Eq.~(\ref{rhs-complex}),
one sees that Fourier coefficients $\hat{f}_q$ with $|q|>2$ are a priori
involved, but they can actually be discarded as being of order
higher than $\epsilon^2$, whereas $\hat{f}_2=\mathcal{O}(\epsilon^2)$.
Similarly, the quantity $Q_l^{(2)}$ can be expressed as a function of
$\hat{f}_1$ and $\hat{f}_3$, and here again, $\hat{f}_3$ can be neglected.
One thus ends up with the following equation for $\hat{f}_2$:
\bea \label{eq-f2}
\nonumber
&& \frac{\partial \hat{f}_2}{\partial t}
+ \frac{v_0}{2} \frac{\partial\hat{f}_1}{\partial x_1}
- \frac{v_0}{2i} \frac{\partial\hat{f}_1}{\partial x_2}
= \\ \nonumber
&& \qquad \qquad
- \left[\lambda \left( 1-e^{-2\sigma_0^2} \right)
+\frac{16}{3\pi}d_0 v_0 \left(\frac{7}{5}+e^{-2\sigma^2}\right)\rho \right]\hat{f}_2 \\
&& \qquad \qquad
+ \frac{8}{\pi}d_0 v_0 \left(\frac{1}{3}+e^{-2\sigma^2}\right) \hat{f}_1^2.
\eea
Within our hydrodynamic description, it is also natural to assume that
the phase-space probability density $f(\mathbf{r},\theta,t)$, or
equivalently, its Fourier coefficients $\hat{f}_k(\mathbf{r},t)$,
vary significantly only over time and length scales that are much larger
than the microscopic ones. Relevant microscopic time scales are the
typical collision time $\tau_\mathrm{col}=1/(\rho d_0 v_0)$,
and the typical ballistic time $\tau_\mathrm{bal}=1/\lambda$
between self-diffusion events.
It is thus legitimate to neglect the term
$\partial \hat{f}_2/\partial t$ in Eq.~(\ref{eq-f2}),
as it is much smaller than $\hat{f}_2/\tau_\mathrm{col}$
and $\hat{f}_2/\tau_\mathrm{bal}$.
In contrast, the terms containing the spatial derivatives have to be
retained, as they involve $\hat{f}_1$ which is much larger than $\hat{f}_2$.

From Eq.~(\ref{eq-f2}) --without the time-derivative term-- one can
express $\hat{f}_2$ as a function of $\rho$ and $\hat{f}_1$. Then
plugging this expression for $\hat{f}_2$ into
Eq.~(\ref{complex-boltz}), with $k=1$, leads to a closed hydrodynamic
equation governing the evolution of $\hat{f}_1$, and involving only
$\hat{f}_1$ and $\rho$.  Mapping back complex numbers onto
two-dimensional vectors, $v_0 \hat{f}_1$ can be identified with the
``momentum'' field $\mathbf{w}=\rho\mathbf{u}$, and one obtains the
following hydrodynamic equation:
\bea \nonumber
 \frac{\partial \mathbf{w}}{\partial t} + \gamma (\mathbf{w} \cdot \nabla)
\mathbf{w} = &-&\frac{v_0^2}{2}\nabla \rho +
\frac{\kappa}{2}\nabla \mathbf{w}^2
+ (\mu - \xi \mathbf{w}^2) \mathbf{w} + \nu \nabla^2\mathbf{w}\\ 
&-& \kappa (\nabla \cdot \mathbf{w}) \mathbf{w}
+2\nu'\nabla\rho\cdot\mathbf{M} - \nu' (\nabla\cdot\mathbf{w})\nabla \rho,
\label{eq-w}
\eea
with $\nu'=\partial\nu/\partial \rho$, and
where $\mathbf{M}=\frac{1}{2}(\nabla\mathbf{w}+\nabla\mathbf{w}^{\mathbf{T}})$
is the symmetric part of the momentum gradient tensor.
The different coefficients appearing in this equation are given by:
\bea 
\nu &=& \frac{v_0^2}{4} \left[ \lambda \left(1-e^{-2\sigma_0^2}\right) 
+\frac{16}{3\pi}d_0 v_0 \rho \left(\frac{7}{5}+
e^{-2\sigma^2} \right)\right]^{-1},\label {eq-nu} \\ 
\gamma &=& \frac{16\nu d_0}{\pi v_0}\left(\frac{16}{15}+
2e^{-2\sigma^2} -e^{-\sigma^2/2} \right),\label{eq-gamma}\\ 
\kappa&=& \frac{16\nu d_0}{\pi v_0} \left(\frac{4}{15}+2e^{-2\sigma^2}
+e^{-\sigma^2/2} \right),\label{eq-kappa}\\
\mu &=& \frac{8}{\pi}d_0 v_0 \rho \left(e^{-\sigma^2/2}-
\frac{2}{3}\right)-\lambda 
\left( 1-e^{-\sigma_0^2/2} \right),\label{eq-mu}\\
\xi &=& \frac{256\nu d_0^2}{\pi^2 v_0^2} \left( e^{-\sigma^2/2}-
\frac{2}{5} \right)\left( \frac{1}{3}+e^{-2\sigma^2} \right).\label{eq-xi}
\eea
Eq.~(\ref{eq-w}) may be considered as a generalization of the
Navier-Stokes equation to a case where on the one hand, the global
momentum of the assembly of particles is not conserved by the
microscopic dynamics, and on the other hand, the dynamics breaks the
Galilean invariance.  This shows up in the appearance of new terms in
the equation, as well as in the presence of the coefficient $\gamma$,
generically different from the Navier-Stokes value $1/\rho$, in front
of the $(\mathbf{w}\cdot\nabla)\mathbf{w}$ term.  For instance, if
$\lambda \ll \rho d_0 v_0$, $\gamma \rho$ remains close to $0.6$ for
any value of $\sigma$.

The different terms in the r.h.s.~of Eq.~(\ref{eq-w}) may be
interpreted as follows. 
Neglecting the density dependence of $\kappa$,
the first two terms can be considered as a pressure
gradient, where the effective pressure $P_\mathrm{eff}$
obeys the equation of state
$P_\mathrm{eff}=\frac{1}{2}(v_0^2 \rho-\kappa \mathbf{w}^2)$.
The third term
accounts for the local relaxation of the momentum field $\mathbf{w}$,
and this term plays an important role in the onset of a collective
behaviour, as we shall see in the following section (note that $\xi>0$
when $\mu>0$).  The fourth term describes the viscous damping, like in
the usual Navier-Stokes equation. The parameter $\nu$ can thus be
interpreted as a kinematic viscosity. It decreases when $\rho$
increases, but the 'dynamic' viscosity $\rho\nu$ increases with
$\rho$. The fifth term may be thought of as a nonlinear
feedback on the momentum field of the compressibility of the flow.
Finally, the two last terms correspond to a coupling between the
density and momentum gradients.

It is also important to note that the above hydrodynamic equation
(\ref{eq-w}) is consistent with the phenomenological equation
postulated by Toner and Tu on the basis of symmetry
considerations~\cite{Toner1995}.  Specifically, expanding the
expression of $\mathbf{w}=\rho\mathbf{u}$ in that equation, we find
the same terms involving the velocity gradients as in
Ref.~\cite{Toner1995}. But it turns out that the term $\nabla(\nabla
\cdot \mathbf w)$, that would be allowed from symmetry considerations,
does not appear in the present approach, that is, the coefficient in
front of it vanishes.  Note also that the term
$(\mathbf{u}\cdot\nabla)^2 \mathbf{u}$ considered by Toner and Tu
\cite{Toner1995}, does not appear here for being of higher order than
the terms retained in the expansion.  Last, hydrodynamic equations
which have been derived through the kinetic approach are entirely
deterministic, while Toner and Tu studied stochastic equations.
However some additional terms also appear, like the coupling terms between
density and velocity gradients.
Most importantly, the present approach provides
a microscopic justification to the hydrodynamic equation of motion,
and yields explicit expressions, as a function of the microscopic
parameters, for the different coefficients appearing in the equations
(transport coefficients).

%%%%%%%%%%%%%%%%%%%%%%%%%%%%%%%%%%%%%%%%%%%%%%%%%%%%%%%%%%%%%%%%%%%

\section{Noise-density phase diagram from the hydrodynamic equations}
\label{sec-phase-diag}

%%%%%%%%%%%
\subsection{Spatially homogeneous stationary solutions}
\label{sec-statsol}

\subsubsection{Transition toward collective motion.}

Now that the hydrodynamic equations of motion have been derived, it is
natural to look for the different possible stationary solutions and to
test their stability. Let us first look for the spatially homogeneous
stationary solutions. Dropping all space and time derivatives, one
ends up with the simple equation:
\be \label{eq-homogstat}
(\mu - \xi \mathbf{w}^2) \mathbf{w} = 0.
\ee
Hence a trivial homogeneous stationary solution is $\mathbf{w}=0$
for all values of the parameters.
When $\mu>0$, a second solution appears, namely
$\mathbf{w}=\mathbf{w}_1=\sqrt{\mu/\xi}\, \mathbf{e}$, where 
$\mathbf{e}$ is a unit vector pointing in an arbitrary direction.
The stability against spatially homogeneous perturbations is easily
tested by assuming that the flow is homogeneous, but time-dependent
in Eq.~(\ref{eq-w}), yielding:
\be \label{eq-homog}
\frac{\partial \mathbf{w}}{\partial t} = (\mu - \xi \mathbf{w}^2) \mathbf{w}.
\ee

\begin{figure}[t]
\centering\includegraphics[width=0.75\textwidth,clip]{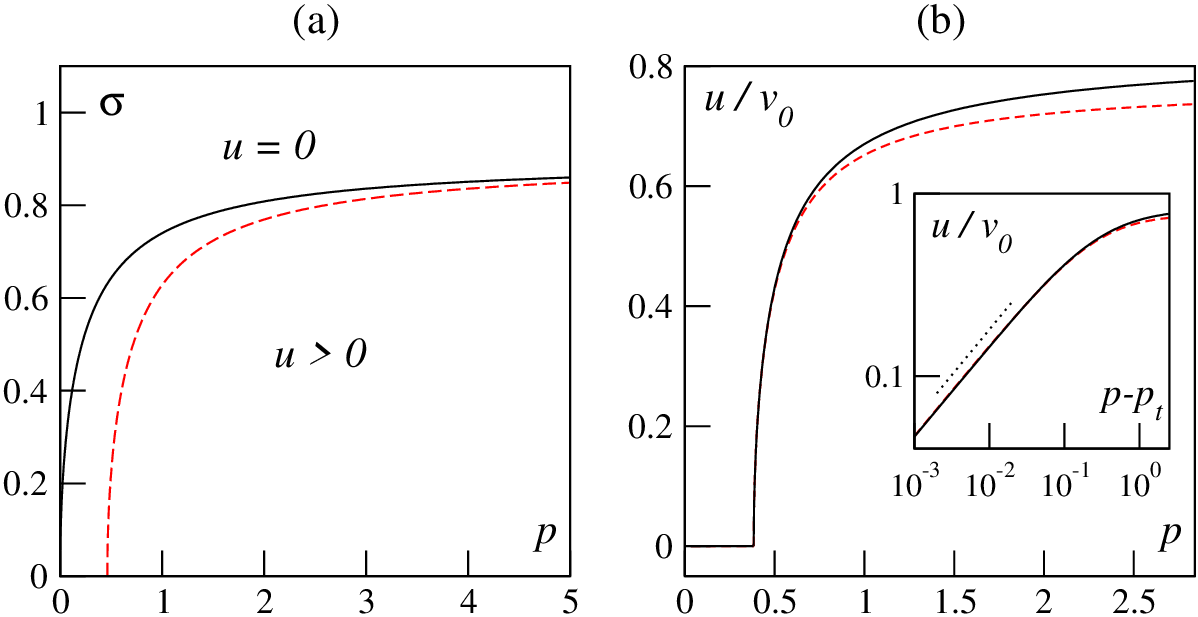}
\caption{(a) Phase diagram of the model in the plane $(p,\sigma)$,
with $p=\rho v_0 d_0/\lambda$.
A transition line (full line: $\sigma_0=\sigma$; dashed line: $\sigma_0=1$)
indicates the linear instability threshold of the state
$u=|\mathbf{u}|=0$.
(b) Hydrodynamic velocity $u$ in the homogeneous state
for $\sigma=\sigma_0=0.6$, computed numerically
from the Boltzmann equation (full line) and analytically from the hydrodynamic
equations (dashed line). Inset: same data on logarithmic scales (dots: slope
$1/2$).
}
\label{phase-diag}
\end{figure}

\noindent
It follows that $\mathbf{w}=0$ is a stable solution when $\mu<0$,
while it becomes unstable for $\mu>0$. In the latter case, the emerging
solution $\mathbf{w}=\mathbf{w}_1$ is stable against homogeneous
perturbations.
From the expression (\ref{eq-mu}) of $\mu$, we see that the sign of $\mu$
is related to a competition between density and self-diffusion.
When the self-diffusion probability $\lambda$ is high, $\mu<0$
and there is no flow. In constrast, when the density is high,
$\mu>0$ and a spontaneous flow appears, due to the numerous
interactions between particles.
The value $\mu=0$ defines a transition line in the phase
diagram noise versus density: for given values $\sigma$ and $\sigma_0$
of the noises, the nonzero solution $\mathbf{w}=\mathbf{w}_1$ appears for
a density $\rho>\rho_\mathrm{t}$, where the threshold density $\rho_\mathrm{t}$ is given by:
\be
\rho_\mathrm{t} = \frac{\pi\lambda(1-e^{-\sigma_0^2/2})}{8d_0 v_0(e^{-\sigma^2/2}-
\frac{2}{3})}.
\label{rho-t}
\ee
In terms of the dimensionless parameter (or reduced density)
\be
p = \frac{B}{H^2} = \frac{d_0\ell_\mathrm{bal} }{\ell_\mathrm{pp}^2}
= \frac{\rho d_0 v_0}{\lambda},
\ee
the threshold is expressed as
\be
p_\mathrm{t} = \frac{\pi(1-e^{-\sigma_0^2/2})}{8(e^{-\sigma^2/2}-
\frac{2}{3})}.
\label{p-t}
\ee
This last result is interesting, as it shows that the threshold $p_\mathrm{t}$,
which could a priori depend on the three dimensionless numbers
$\sigma$, $\sigma_0$ and $B$, actually does not depend on $B$.
The transition line is plotted in Fig.~\ref{phase-diag}(a) for the two
cases $\sigma_0=\sigma$ and $\sigma_0=1$.  Instead of considering the
transition as a function of the density, one may also look for the
transition by varying the noises at a given fixed density.  If the two
noise intensities $\sigma_0$ and $\sigma$ are equal, the instability of
$\mathbf{w}=0$ occurs for any (non-zero) density, and the threshold
noise $\sigma_\mathrm{t}$ behaves in the low density limit $p \to 0$ as
$\sigma_\mathrm{t} \sim p^{1/2}$.  This nontrivial prediction can be
verified in direct numerical simulations (see below). In contrast,
when $\sigma_0$ is kept fixed while varying $\sigma$, no transition
occurs as a function of $\sigma$ if the reduced density is lower than
a limit $p_\mathrm{t}^0$ given by:
\be
p_\mathrm{t}^0 = \frac{3\pi}{8}(1-e^{-\sigma_0^2/2}).
\ee
Finally, in the opposite limit of
high density, the threshold noise $\sigma_\mathrm{t}$ saturates to a value
$\sigma_\mathrm{t}^{\infty}=(2\ln\frac{3}{2})^{1/2} \approx 0.90$.

\subsubsection{Validity domain of the hydrodynamic equations.}
\label{sec-valid}

The hydrodynamic equations rely on the scaling ansatz (\ref{eq-scal-ansatz}).
In order to verify a posteriori the validity of the hydrodynamic equations,
we compare the stationary homogeneous solutions with non-zero velocity
obtained from the hydrodynamic equations to that numerically computed from the
Boltzmann equation.
The hydrodynamic velocity $u$, computed as
$u = u_1 \equiv \rho^{-1} \sqrt{\mu/\xi}$,
is plotted on Fig.~\ref{phase-diag}(b)
as a function of the reduced density $p$.
Note that $u/v_0$ is a function of the dimensionless numbers $p$, $\sigma$
and $\sigma_0$ only. As expected, the velocity
$u$ computed from the hydrodynamic equation
matches perfectly, in the small velocity regime (i.e., close to
the transition line) the numerical data from the Boltzmann equation.
However, it turns out that even quite far from the transition, when
$u$ becomes of the order of $v_0$, the value $u_1$ computed from
the hydrodynamic equation remains a good estimate of the value obtained
from the Boltzmann equation. In particular, it is interesting to note
that $u_1$ also saturates at large $\rho$ to a finite value
$u_1^{\infty}(\sigma)<1$, given by
\be
u_1^{\infty}(\sigma) = v_0 \left[\frac{2\left(e^{-\sigma^2/2}-\frac{2}{3}
\right) \left(\frac{7}{15}+\frac{1}{3}\, e^{-2\sigma^2}\right)}
{\left(e^{-\sigma^2/2}-\frac{2}{5}\right)
\left(\frac{1}{3}+ e^{-2\sigma^2}\right)} \right]^{\frac{1}{2}}
\ee
(see Fig.~\ref{phase-diag}(b)).
Hence, even beyond their domain of validity, which is restricted to
small values of the hydrodynamic velocity, the hydrodynamic
equations we have derived yield a rather good approximation of the
exact dynamics. Specifically, they fulfill the condition that the
hydrodynamic velocity should remain smaller than the individual velocity
$v_0$ of the particles, although this result was not a priori obvious
given the approximations made.

\begin{figure}[t]
\psfrag{G1}{$g_k$}
\psfrag{G2}{$g_k/g_1^k$}
\centering\includegraphics[width=0.75\textwidth,clip]{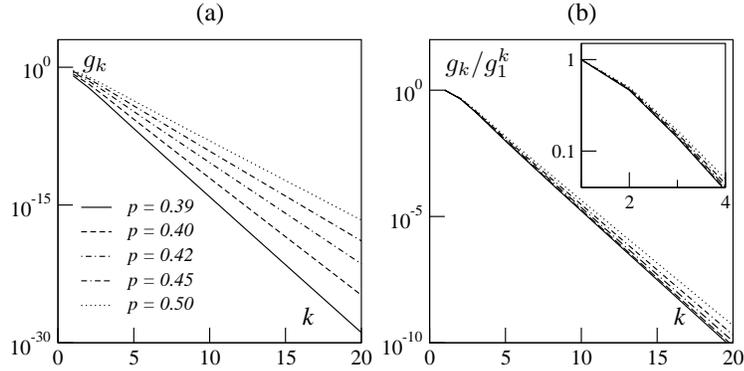}
\caption{Test of the scaling ansatz
  $\hat{f}_k=\mathcal{O}(\epsilon^{|k|})$. (a)
  $g_k=\hat{f}_k^\mathrm{st}/\rho$ versus $k$, for
  $\sigma=\sigma _0=0.6$ and different values of the reduced density
  $p$, close to the transition density ($p_\mathrm{t}=0.3837$); an exponential
  decay is observed. (b) $g_k/g_1^k$ as a function of $k$, showing
  that for a given $k$, $g_k$ is essentially proportional to $g_1^k$
  when the density is varied ($\lambda=0.5$, $d_0=0.5$,
  $v_0=1$). Inset: zoom on the small $k$ region.
}
\label{scal-ansatz}
\end{figure}

To further test the validity of the hydrodynamic equations,
we have also checked explicitely, from a numerical calculation,
that the scaling ansatz (\ref{eq-scal-ansatz}) is correct.
Specifically, we computed from a numerical integration the stationary
and spatially homogeneous solution $\hat{f}_k^\mathrm{st}$ of the
Boltzmann equation.
In order to work with dimensionless quantities, we plot on
Fig.~\ref{scal-ansatz}(a) the quantities
$g_k=\hat{f}_k^\mathrm{st}/\rho$ (instead of $\hat{f}_k^\mathrm{st}$)
as a function of $k$.
We observe that $g_k$ decays almost exponentially with $k$, as soon as
$k \gtrsim 4$. To test the scaling ansatz, we first reformulate it in
a more specific way. The ansatz is obeyed if there exists for all $k$ a
constant $c_k$ such that $g_k \approx c_k g_1^k$ in a parameter regime
where $g_1 \ll 1$.
We thus plot on Fig.~\ref{scal-ansatz}(b) the ratio $g_k/g_1^k$ for
different values of the density, close to the transition, and we observe
a reasonable collapse of the data.  Let us however emphasize that a
strict collapse of the data is not necessary in order to apply the
approximation scheme used in the derivation of the hydrodynamic
equations. The essential requirement is that the quantities $g_k$ with
$k>2$ could be neglected. As the ratio $g_k/g_1^k$ decays rapidly with
$k$, neglecting terms with $k>2$ is a safe approximation.

%%%%%%%%%%%
\subsection{Stability against inhomogeneous perturbations of the homogeneous
stationary solutions}
\label{subsect-stab-analysis}

\subsubsection{Evolution equation for the perturbations.}

We have shown that above a threshold density
$\rho_\mathrm{t}$, or equivalently, below a threshold noise $\sigma_\mathrm{t}$,
the solution with zero velocity becomes unstable, and a stable solution
with finite velocity emerges. Yet, only the stability with respect to
homogeneous perturbations (i.e., with infinite wavelength) has been tested
up to now. Hence this does not ensure that the finite velocity solution is
really stable, as it may be destabilized by finite wavelength perturbations.
We now check this issue, by introducing small perturbations around
the homogeneous stationary solutions $\rho_0$ and $\mathbf{w}_0$, namely
\be \label{perturb}
\rho(\mathbf{r},t) = \rho_0 + \delta \rho(\mathbf{r},t),
\quad
\mathbf{w}(\mathbf{r},t) = \mathbf{w}_0 + \delta \mathbf{w}(\mathbf{r},t).
\ee
Note that $\mathbf{w}_0$ may either be equal to zero or to the nonzero
solution $\mathbf{w}_1$.
Plugging these expressions into the hydrodynamic equations
(\ref{continuity}) and (\ref{eq-w}),
we can expand the resulting equations to first order in the perturbation
fields $\delta \rho(\mathbf{r},t)$ and $\delta \mathbf{w}(\mathbf{r},t)$,
also taking into account the density dependence of the different coefficients.
This yields the following linearized equations:
\bea \label{lin-rho}
&&\frac{\partial}{\partial t} \delta \rho + \nabla \cdot
\delta \mathbf{w} = 0,\\
\label{lin-w}
&&\frac{\partial}{\partial t} \delta \mathbf{w}
+\gamma(\mathbf{w}_0\cdot\nabla)\delta \mathbf{w} =
-\frac{v_0^2}{2} \nabla\delta \rho
+\kappa\nabla (\mathbf{w}_0 \cdot \delta\mathbf{w}) \\ \nonumber
&& \qquad \qquad \qquad +[(\mu'-\xi' \mathbf{w}_0^2) \delta\rho
-2\xi\mathbf{w}_0 \cdot \delta \mathbf{w}
- \kappa \nabla\cdot\delta \mathbf{w}] \mathbf{w}_0\\ \nonumber
&& \qquad \qquad \qquad +(\mu-\xi\mathbf{w}_0^2)\delta \mathbf{w}
+\nu \nabla^2 \delta \mathbf{w},
\eea
where $\mu'$ and $\xi'$ are shorthand notations for
$\partial \mu/\partial \rho$ and
$\partial \xi/\partial \rho$.  Note that $\partial \mu/\partial \rho$
is actually a constant, i.e., it is independent of $\rho$.
Then we make the following ansatz \be
\delta \rho(\mathbf{r},t) = \delta \rho_0\,
e^{st+i\mathbf{q}\cdot\mathbf{r}}, \quad \delta
\mathbf{w}(\mathbf{r},t) = \delta \mathbf{w}_0 \,
e^{st+i\mathbf{q}\cdot\mathbf{r}}, \ee where $\delta \mathbf{w}_0$ is
a vector (with real components), and $\delta \rho_0$ is a complex
amplitude that takes into account a possible phase shift between
density and momentum perturbation fields.  Both $||\delta
\mathbf{w}_0||$ and $|\delta \rho_0|$ are assumed to be small.  The
wavenumber $\mathbf{q}$ is assumed to have real components, whereas
the growth rate $s$ is a priori complex.  In addition, $\mathbf{q}$ is
considered to be given, and one looks for the dispersion relation
$s(\mathbf{q})$. If the real part $\Re[s(\mathbf{q})]>0$, the
mode with wavenumber $\mathbf{q}$ is unstable.  Then
Eqs.~(\ref{lin-rho}) and (\ref{lin-w}) become:
\bea \label{lin-rho-sq}
&& s\, \delta\rho_0 + i\mathbf{q}\cdot\delta\mathbf{w}_0 =
0,\\ \nonumber && [s+\gamma(\mathbf{w}_0\cdot
  i\mathbf{q})-(\mu-\xi\mathbf{w}_0^2)+ \nu\mathbf{q}^2] \delta
\mathbf{w}_0 =\\ \nonumber && \qquad \qquad
-\frac{1}{2}(v_0^2 \, \delta\rho_0
  -2\kappa\mathbf{w}_0\cdot\delta\mathbf{w}_0) i\mathbf{q} \\ && \qquad
\qquad +[(\mu'-\xi'\mathbf{w}_0^2)\delta\rho_0
  -(2\xi\mathbf{w}_0+ \kappa i\mathbf{q})\cdot
  \delta\mathbf{w}_0]\mathbf{w}_0.
\label{lin-w-sq}
\eea
Note that, due to linearity, the above equations can be re-expressed as
a function of the ratio of amplitudes $\delta\mathbf{w}_0/\delta\rho_0$.

\subsubsection{Stability of the zero-velocity solution.}

Let us first check the stability against inhomogeneous perturbations
of the solution $\mathbf{w}_0=0$, which is known to be stable
against homogeneous perturbations in the low density
phase $\rho<\rho_\mathrm{t}$ (corresponding to $\mu<0$).
In this case, Eq.~(\ref{lin-w-sq}) simplifies to:
\be
(s+\nu \mathbf{q}^2-\mu)\delta \mathbf{w}_0 = -\frac{i}{2} v_0^2
\delta \rho_0\mathbf{q}.
\ee
Thus $\delta \mathbf{w}_0$ is along the same direction as $\mathbf{q}$.
Writing $\mathbf{q}=q\mathbf{e}$ and $\delta \mathbf{w}_0=\delta w_0
\mathbf{e}$, where $\mathbf{e}$ is an arbitrary unit vector, one can
eliminate the ratio $\delta w_0/\delta \rho_0$ from Eq.~(\ref{lin-rho-sq}),
yielding:
\be
s^2+(\nu q^2-\mu)s+\frac{v_0^2}{2}q^2=0.
\ee
The discriminant of this second order polynomial equation
reads (note that $\mu<0$):
\be
\Delta = (|\mu|+\nu q^2)^2 -2v_0^2q^2.
\ee
If $\Delta \ge 0$, the roots are real, and one finds for the largest one
$s_+$:
\be
s_+=\frac{1}{2} \left[-(|\mu|+\nu q^2)
+\sqrt{(|\mu|+\nu q^2)^2-2v_0^2q^2}\right] <0.
\ee
In the opposite case $\Delta<0$, the roots $s_{\pm}$ are complex
conjugates, and their real part is given by:
\be
\Re[s_{\pm}]=-\frac{1}{2} (|\mu|+\nu q^2)<0.
\ee
As a consequence, the homogeneous fields $\mathbf{w}_0=0$ is stable with
respect to finite wavelength perturbations in the region $\rho<\rho_\mathrm{t}$.

\subsubsection{Stability of homogeneous collective motion.}
\label{stab-hmg}

We now turn to the stability analysis of the stationary homogeneous flow
$\mathbf{w}_0=\mathbf{w}_1$, obtained for $p>p_\mathrm{t}$.
For the hydrodynamic equations to be valid, we restrict our study to
values of $p$ very close to $p_\mathrm{t}$, with $p>p_\mathrm{t}$.
One could a priori consider vectors $\mathbf{q}$ and $\delta\mathbf{w}_0$
that make arbitrary angles with respect to $\mathbf{w}_1$.
However, it can be shown (see \ref{app-stability}) that only some specific
angles are allowed. Further, for all allowed perturbation modes such that
$\mathbf{q}$ and $\delta\mathbf{w}_0$ are not along the direction
of $\mathbf{w}_1$, the real part of the growth rate $s$ is negative,
so that these modes are stable (\ref{app-stability}).
The only instability that appears is for longitudinal perturbations,
such that $\mathbf{q}$, $\delta\mathbf{w}_0$ and $\mathbf{w}_1$ all have
the same direction. We thus focus on this specific case in the following.

\begin{figure}
\begin{center}
\psfrag{Re[s]}{$\Re[s_+]/\lambda$}
\includegraphics[width=0.75\textwidth,clip]{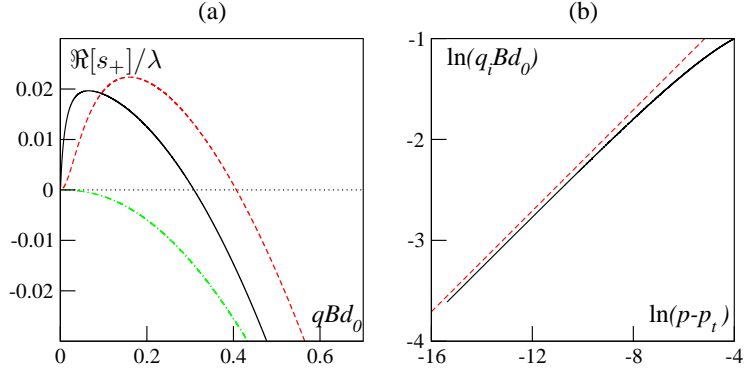}
\caption{
Longitudinal instability. (a) $\Re [s_+]/\lambda$
  \emph{versus} $q$ for $p=0.22$ (full line), $0.30$ (dashed) and
  $p=0.4$ (dot-dashed), and $\sigma=\sigma _0=0.5$
  ($p_\mathrm{t} = 0.2138$). The maximum growth rate decreases when $p$ is
  increased. (b) $\left(q_iBd_0\right)$ \emph{vs} $(p-p_\mathrm{t})$ in
  logarithmic scales. The dashed line indicates the scaling
  $q_i\propto(p-p_\mathrm{t})^{1/4}$.}
\label{Fdispersion1}
\end{center}
\end{figure}

Considering a longitudinal perturbation, we write $\mathbf{w}_1=w_1\mathbf{e}$,
$\mathbf{q}=q\mathbf{e}$ and $\delta\mathbf{w}_0=\delta
w_0\mathbf{e}$, where $\mathbf{e}$ is a unit vector.  Under these
assumptions, Eqs.~(\ref{lin-rho-sq}) and (\ref{lin-w-sq}) become:
\bea
\label{lin-rho1-sq}
&& s\,\delta\rho_0+iq\,\delta w_0=0, \\
\label{lin-w1-sq}
&& (s+\gamma iqw_1+\nu q^2)\delta w_0 = -\frac{iq}{2}(v_0^2
\delta \rho_0 -2\kappa w_1 \delta w_0)\\ \nonumber
&& \qquad \qquad + w_1[(\mu'-\xi' w_1^2)\delta \rho_0 -
(2\xi w_1+iq\kappa)\delta w_0],
\eea
where we also take into account that $\mu-\xi w_1^2=0$.
From Eq.~(\ref{lin-rho1-sq}), one gets $\delta w_0/\delta\rho_0=-s/iq$,
which we report in Eq.~(\ref{lin-w1-sq}). This yields a polynomial
of second degree in $s$
\bea
s^2&+&s\left[\left(\nu q^2+2\mu\right)+iq\gamma w_1\right]
\\
&+&\left[\frac{q^2 v_0^2}{2}
+iqw_1\left(\mu'-\xi' w_1^2\right)\right]=0,\nonumber
\eea
from which two solutions $s_{\pm}$ can be obtained.
Denoting as $s_+$ the solution with the largest real part, we find
\be
\Re [s_+]=-\frac{1}{2} \left(\nu q^2+2\mu \right)
+\sqrt{\frac{1}{8}\left(J_1+\sqrt{J_1^2+J_2^2}\right)},
\label{Eq_Res}
\ee
with
\bea
J_1 &=& \left(\nu q^2+2\mu\right)^2-q^2\left(\gamma^2w_1^2+
2v_0^2\right)\nonumber\\
J_2 &=& 2w_1q [\gamma (\nu q^2+2\mu)-2\mu'+2\xi' w_1^2].
\nonumber
\eea
To deal with this complicated expression, we first plot $\Re [s_+]$
as a function of $q$ for some specific values of the parameters
(see Fig.~\ref{Fdispersion1}(a)). 
Near the threshold $p_\mathrm{t}$ of collective motion, there exists a
threshold value $q_i$ such that $\Re [s_+]$ is positive for 
$q<q_i$ and negative for $q>q_i$. Hence the homegeneous flow
turns out to be unstable with respect to long wavelength perturbations.

This result is confirmed by a small $q$ expansion of Eq.~(\ref{Eq_Res}).
Expanding $\Re[s_+]$ up to second non-trivial order in $q$, that is
to order $q^4$ since only even powers of $iq$ appear in the expansion
of the real part of $s_{+}$, we get\footnote{To simplify the resulting
expressions, we approximate the coefficients of the expansion in
$q$ by their leading order in $1/\mu$, as $\mu$ is small close to the
transition line. The full expression of the coefficient $s_2$ of the
$q^2$ term reads:
$$
s_2 = \frac{1}{8} \left[ \frac{1}{\xi} \left( \frac{\mu'}{\mu}
- \frac{\xi'}{\xi} -\gamma \right)^2 -\frac{\gamma^2}{\xi}
-\frac{2 v_0^2}{\mu} \right].
$$
This expression will be used in Fig.~\ref{fig-s2} to compare with
numerical results.
}.
\be \label{re-s}
\Re[s_+] = \frac{\mu'^2}{8\,\xi \mu^2} \, q^2 -
\frac{5\,\mu'^4}{128\, \xi^2 \mu^5} \, q^4
+ \mathcal{O}(q^6).
\ee
The positivity of the coefficient of the $q^2$ term confirms that,
close to the transition line, long wavelength modes are unstable.
Note that the expansion (\ref{re-s}) is consistent as long as the fourth order
term remains small with respect to the second order one,
yielding the condition $q \ll q^*$ which defines the wavenumber $q^*$:
\be
q^* = d_0^{-1} B^{-1} \left(p-p_\mathrm{t}\right)^{\frac{3}{2}} \Psi(p).
\label{eq-qetoile}
\ee
The function $\Psi(p)$ goes to a constant value for $p\to 0$, and
$\Psi(p) \sim p^{-1/2}$ for $p \to \infty$.
The wave vector $q^*$ defines the
region where the first term of the expansion of $\Re[s_+]$ is
dominant. 

Interestingly, we observe that other wavenumbers characterizing
$\Re[s_+]$ have a different scaling with $p-p_\mathrm{t}$,
the deviation from the threshold.
For instance, it can be shown analytically that the wavenumber $q_i$,
defined by $\Re[s_+]=0$, scales as $q_i \sim w_1^{1/2} \sim (p-p_\mathrm{t})^{1/4}$,
as illustrated on Fig.~\ref{Fdispersion1}(b).
The wavenumber $q_i$ delimitates the domain of unstable modes.
Another example is given by the wavenumber $q_m$ that maximizes
$\Re[s_+]$, and thus corresponds to the most unstable modes:
$q_m$ is found to scale as $q_m \sim w_1 \sim (p-p_\mathrm{t})^{1/2}$.
The existence of these different scaling regimes is an illustration
of the complexity of the dynamics close to the transition line.

Finally, we emphasize that the
perturbations that destabilize the homogeneous collective motion (that
is, the long-range order) are different from the ones that destabilize
long-range order in the XY-model, an equilibrium model with
essentially the same symmetries as in the present model. In our model,
motion is destabilized by longitudinal waves, while in the XY-model,
long-range order is destabilized by spin-waves, that is, by a small
change in the spin direction from one spin to the neighbouring ones.

%%%%%%%%%%%
\subsection{Comparison with the phase diagram of the agent-based model}

\subsubsection{Numerics and parameters.}

All simulations are performed using models defined on a square domain,
with periodic conditions on both boundaries.
The initial conditions always consist in randomly dispersed
particles, with a uniformly chosen random speed direction. Then all
measurements are performed after a sufficiently long time so that a
stationary state is reached. 

\begin{table}[b]
  \begin{center}
    \begin{tabular}{|c|c|c|c|c|c||c|c|}\hline
      set of parameters& $\sigma _0$&$d_0$&$v_0$&$\lambda$&$\rho$&$p$&B\\\hline
  I   &$\sigma$&$1$&$2^{-1}$&$1$&$[2^{-7};2^2]$&$[2^{-6};2]$&$2^{-1}$\\
  II  &$\sigma$&$1$&$2^{-1}$&$2^{-3}$&$[2^{-9};2^{-4}]$&$[2^{-7};2^{-2}]$&$2^2$\\
  III &$\sigma$&$1$&$2^{-1}$&$2^{-4}$&$[2^{-10};2^{-4}]$&$[2^{-7};2^{-1}]$&$2^3$\\
  IV  &$\sigma$&$1$&$2^{-1}$&$2^{-5}$&$[2^{-11};2^{-4}]$&$[2^{-7};1]$&$2^4$\\
  V   &$\sigma$&$1$&$1$&$2^{-5}$&$[2^{-10};2^{-6}]$&$[2^{-7};2^{-3}]$&$2^5$\\
  VI  &$\sigma$&$2^{-1}$&$2^{-1}$&$2^{-5}$&$[2^{-11};2^{-5}]$&$[2^{-8};2^{-2}]$&$2^5$\\
  VII &$\sigma$&$2^{-2}$&$2^{-1}$&$2^{-5}$&$[2^{-11};2^{-4}]$&$[2^{-9};2^{-2}]$&$2^6$\\
  VIII&$\sigma_0^\mathrm{max}$&$2^{-2}$&$2^{-1}$&$2^{-5}$&$[2^{-12};2^{-5}]$&$[2^{-10};2^{-2}]$&$2^6$\\ 
\hline
    \end{tabular}
    \caption{Physical parameters of agent-based
      simulations; set I corresponds to the values used in~\cite{Chate2008}.
      Parameters are chosen as multiple or sub-multiple of $2$;
      $\sigma_0^\mathrm{max}=\pi/\sqrt{3}$ corresponds
      to the same variance as a uniform noise on $[-\pi;\pi]$.}
    \label{Tparam}
  \end{center}
\end{table}

In the above framework of the Boltzmann equation, we considered
diluted systems with small correlations between particles, which is
expressed in terms of the dimensionless number $H$ and $B$ as:
\be \nonumber
H \gg 1, \qquad B \gg 1.
\ee
In the numerical agent-based model, we do not have access to very large
values of $H$ and $B$, due to simulation constraints.
However, to be as consistent as possible with the kinetic theory approach,
we mainly explored a parameter range such that $H\ge 4$ and $B\ge 4$.
Among the three dependent dimensionless numbers $B$, $H$ and $p$, we decided to
keep $B$ to characterize the set of parameters, and $p$ as the control
parameter.

Throughout the study, we fix $\Delta t=1$. We defined some sets of parameters
$(d_0,v_0,\lambda)$ and, for each of them, we studied the behaviour
of the system in the parameters space $(\rho,\eta)$. To make
the comparison between analytical and numerical results easier,
we characterize the noise amplitude by its \emph{rms}-value $\sigma$
(or equivalently its variance $\sigma^2$).
For a uniform noise on the interval $[-\eta\pi,\eta\pi]$, we have
$\sigma=\eta.\pi/\sqrt{3}$. The self-diffusion noise is kept equal
to the collision noise ($\sigma_0=\sigma$), except for one
set of parameters in which the angle of diffusion $\eta_0\xi_j^t$ is
chosen over the whole circle. We call its \emph{rms}-value
$\sigma_0^\mathrm{max}$.  All the parameter values are summarised in
Table~\ref{Tparam}.

\subsubsection{Transition line.}
\begin{figure}[t]
\centering\includegraphics[width=0.75\textwidth,clip]{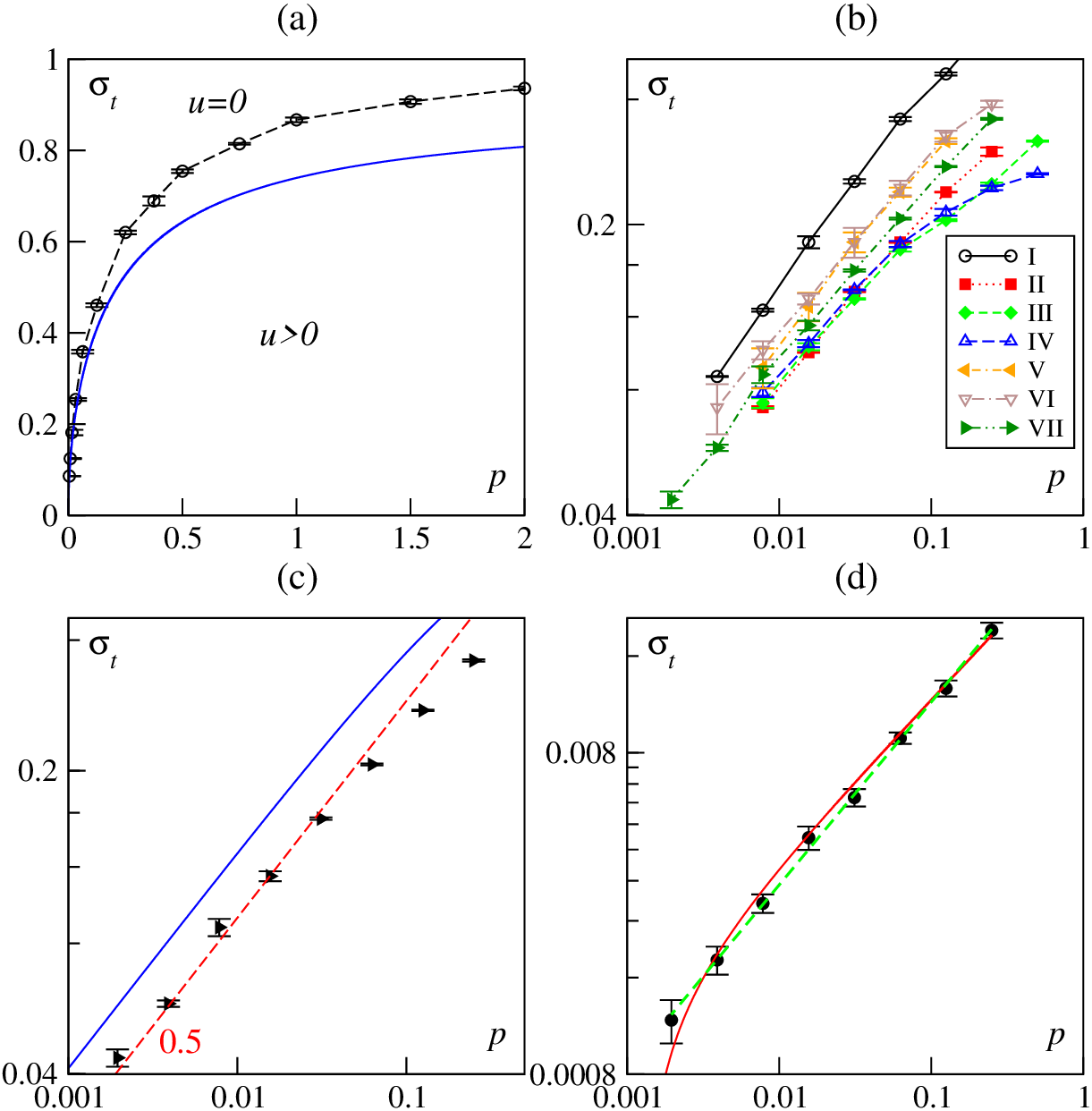}
\caption{Phase diagram of agent-based models. (a) Overview of the
  phase diagram. The continuous line is the transition line of the
  continuous model given in Eq.~(\ref{p-t}); data for symbols $\circ$ are
  obtained with the parameter set I. (b) Diagrams for
  all configurations with the self-diffusion noise $\sigma _0=\sigma$
  (plot in log-log scales). (c) Scaling of the transition line at
  small $p$ with the set of parameters VII ($B=64$) in log-log
  scales. The continuous line is the transition line (\ref{p-t})
  obtained from the kinetic theory.
  The dashed line is a square-root fit of the numerical results.
  (d) Model with a constant and
  maximum noise amplitude for self-diffusion in log-log scales (set
  VIII, $B=64$). The continuous line corresponds to a fit of the
  numerical points with the law $\sigma_\mathrm{t}=\alpha (p-p_\mathrm{t}^0)^{1/2}$. The
  dashed line is a fit with a power law $\beta=0.57$ (see
  Table~\ref{Tparam} for values of the other parameters).}
\label{diag_num}
\end{figure}

When the noise amplitudes for collision and self-diffusion are equal, the
general aspect of the phase diagram is the same both for the kinetic
theory, and the agent-based model (Fig.~\ref{diag_num}(a)). We have drawn
the transition line for different sets of parameters on
Figure~\ref{diag_num}(b). All curves seem to be bounded between
configurations I ($B=0.5$) and II ($B=4$).

Looking at the influence of the different parameters, we can make the
following observations. First, there are small variations as the
self-diffusion probability $\lambda$ changes with a factor of eight from
configuration II to IV (Fig.~\ref{diag_num}(b)). When the dimensionless
parameter $B$ is kept constant (Sets V and VI), the transition
points corresponding to the same value of $p$ are equal
within the error bars. 
Apart from Set I for which $B<1$, it turns out that the measured values of
$\sigma_\mathrm{t}$ differ by less than $15\%$ for any given $p$, while $B$
is varied by a factor of $16$ between Set II ($B=4$) and Set VII ($B=64$).
However, we are not able to conclude that the
curves merge into a single master curve.
In particular, the observed evolution of $\sigma_\mathrm{t}$ when increasing $B$
at fixed $p$ is not monotonous (Fig.~\ref{diag_num}(b)).

In the low $p$ region, the transition noise varies as a power law with
$p$, $\sigma\propto p^\beta$ when $p \to 0$.
We have measured the exponent $\beta$ for the largest dimensionless number
$B$ ($B=64$, set VII), yielding $\beta=0.46\pm0.04$ (Fig.~\ref{diag_num}(c)).
This value is compatible with a square-root behaviour as found analytically
in the binary collision model (Fig.~\ref{phase-diag}).
Quantitatively, the transition line computed from the kinetic approach
and the one which we measure in the agent-based model are relatively
close one to the other; their largest relative difference is about $30\%$.

\subsubsection{Maximal self-diffusion.}
When we set the amplitude of self-diffusion noise to its maximum
($\eta_0=1$ or $\sigma_0=\pi/\sqrt{3}$), the behaviour of the model
remains qualitatively similar to the case $\sigma=\sigma_0$ that
we studied above, with only a few quantitative differences.
The transition line is shifted to a lower noise amplitude: $\sigma_\mathrm{t}$
differs by two orders of magnitude between the two comparable parameter
sets VII and VIII. Fitting the two curves by a power law, the exponents are
significantly different: $\beta\approx 0.46$ for set VII ($\sigma=\sigma_0$),
while $\beta\approx 0.57$ for set VIII ($\sigma_0=\sigma_0^\mathrm{max}$).
One possible explanation for such a difference would be that,
as in the hydrodynamic equations, there exists a threshold $p_\mathrm{t}^0$
below which no collective motion occurs, whatever the noise amplitude
$\sigma$. A fit with the function $p=\alpha\sqrt{p-p_\mathrm{t}^0}$
gives a value $p_\mathrm{t}^0=0.00133$ (Fig.~\ref{diag_num}(d)),
much smaller than the theoretical value $p_\mathrm{t}^0=3\pi/8\approx 1.18$.
Given the presently available data, we are not able to discriminate
between the two fits, and to conclude on the existence of a non-zero
threshold value $p_\mathrm{t}^0$.
Trying to find a phase transition for a very low value of $p$
($p=2^{-10}$), below the fitted value $p_\mathrm{t}^0$,
we could hardly define a threshold. However, it might be necessary
to reach larger system sizes to detect a phase transition in this regime.

%%%%%%%%%%%%%%%%%%%%%%%%%%%%%%%%%%%%%%%%%%%%%%%%%%%%%%%%%%%%%%%%%%%%%%

\section{Beyond the strict validity domain of the hydrodynamic equations}
\label{sec-beyond-valid}

In Section~\ref{sec-phase-diag}, we concluded from a linear stability
analysis that the homogeneous flow is unstable with respect
to long wavelength perturbations, in the validity domain of the
hydrodynamic equations, namely close to the transition line.
When getting farther from the transition line, previous theoretical
approaches \cite{Toner1998} suggest that the homogeneous motion
should be stable. To come to a conclusion in the present framework,
it is thus necessary to come back to an analysis of the Boltzmann
equation. It is also natural to wonder whether the hydrodynamic
equations could yield, out of their strict validity domain, a
qualitative description of the phenomenology of the moving phase. We
address these issues in the present section. We find in particular a
restabilization of the homogeneous flow far from the transition line,
as well as solitary waves that we compare with the travelling stripes
already reported in numerical simulations of the agent-based model
\cite{Gregoire2004}.

\subsection{Stability analysis from the Boltzmann equation}
\label{sec-stab-boltz}

In order to analyse the stability of the finite velocity solution
beyond the validity domain of the hydrodynamic equations, we
come back to the Boltzmann equation, and we resort to a semi-analytical
treatment.

We start with a formal expansion of the phase-space distribution
$f(\mathbf{r},\theta,t)$ around the homogeneous stationary solution
$f_0(\theta)$:
\be
f(\mathbf{r},\theta,t) = f_0(\theta) + \delta f(\mathbf{r},\theta,t).
\ee
Considering a perturbation of wavevector $\mathbf{q}$
of the form:
\be
\delta f(\mathbf{r},\theta,t) = \delta \rho_0\, G(\theta,\mathbf{q})\,
e^{st+i\mathbf{q}\cdot\mathbf{r}},
\ee
with $\int_{-\pi}^{\pi} d\theta\, G(\theta,\mathbf{q})=1$.
Assuming, as in Sect.~\ref{stab-hmg}, that both $\mathbf{q}$ and
the velocity perturbation are along the same direction $\mathbf{e}$
as the collective velocity, the function $G(\theta,\mathbf{q})$
satisfies the following linearized Boltzmann equation:
\be \label{eq-sG}
sG(\theta,\mathbf{q})+iqv_0 \cos\theta\, G(\theta,\mathbf{q})
= I_\mathrm{dif}[G] + I_\mathrm{col}[G,f_0]+I_\mathrm{col}[f_0,G].
\ee
Setting $\mathbf{q}=q\mathbf{e}$,
we are interested in a small $q$ expansion of Eq.~(\ref{eq-sG}), in order to
compare with the results of Eq.~(\ref{re-s}).
We then expand $s$ and $G(\theta,\mathbf{q})$ in the following way:
\bea
&& s = is_1 q+s_2 q^2 + \mathcal{O}(q^3), \\
&& G(\theta,\mathbf{q}) = G_0(\theta)+iq\, G_1(\theta) + q^2\, G_2(\theta) + \mathcal{O}(q^3),
\eea
with the normalization conditions:
\be
\int_{-\pi}^{\pi} d\theta\, G_0(\theta) = 1, \quad
\int_{-\pi}^{\pi} d\theta\, G_1(\theta) = 0, \quad
\int_{-\pi}^{\pi} d\theta\, G_2(\theta) = 0.
\ee
Then $G_0$, $G_1$ and $G_2$ are solutions of the hierarchy of equations:
\bea
\label{eq-G0}
&& I_\mathrm{dif}[G_0] +I_\mathrm{col}[G_0,f_0]+I_\mathrm{col}[f_0,G_0] = 0,\\
\label{eq-G1}
&& s_1 G_0(\theta) + v_0 \cos\theta \,G_0(\theta) = I_\mathrm{dif}[G_1]
+I_\mathrm{col}[G_1,f_0]+I_\mathrm{col}[f_0,G_1],\\
&& -s_1 G_1(\theta)+s_2 G_0(\theta)-v_0 \cos\theta \,G_1(\theta) =\\ \nonumber
&& \qquad \qquad \qquad \qquad I_\mathrm{dif}[G_2]+I_\mathrm{col}[G_2,f_0]+I_\mathrm{col}[f_0,G_2].
\eea
Using the properties of $I_\mathrm{dif}$ and $I_\mathrm{col}$, namely
\be
\!\!\!\!\!\int_{-\pi}^{\pi} d\theta\, I_\mathrm{dif}[g]=0, \quad
\int_{-\pi}^{\pi} d\theta\, I_\mathrm{col}[g,f_0]=
\int_{-\pi}^{\pi} d\theta\, I_\mathrm{col}[f_0,g]=0
\ee
for any function $g$, we obtain
\bea
s_1 &=& -v_0 \int_{-\pi}^{\pi} d\theta \,\cos\theta \, G_0(\theta),\\
s_2 &=& v_0 \int_{-\pi}^{\pi} d\theta \,\cos\theta \, G_1(\theta).
\eea
Hence the determination of $G_2$ is not necessary to compute $s_2$.
We only need to compute the hierarchy of functions up to $G_1$.
It is actually convenient to work in Fourier space, introducing the Fourier
series expansion $\hat{G}_{0,k}$ and $\hat{G}_{1,k}$ of $G_0(\theta)$ and
$G_1(\theta)$ respectively.
In terms of this Fourier expansion, one finds $s_1=-\hat{G}_{0,k=1}$
and $s_2=\hat{G}_{1,k=1}$, assuming that $G_0(\theta)$ and $G_1(\theta)$
are even functions.

\begin{figure}[t]
\centering\includegraphics[width=0.75\textwidth,clip]{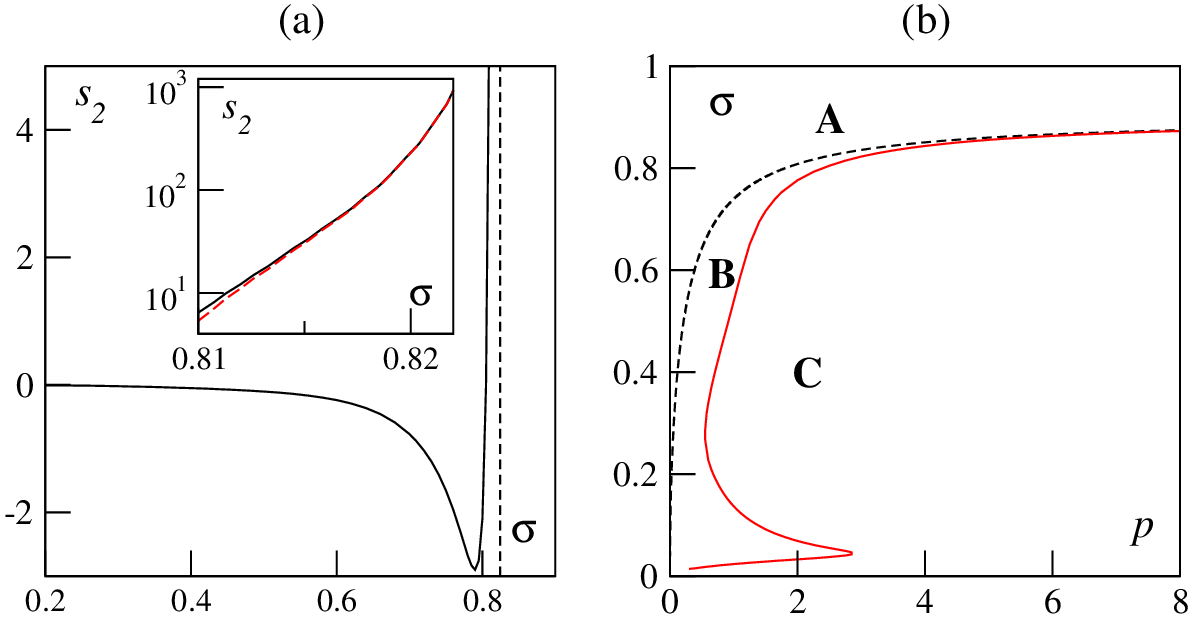}
\caption{(a) Dependence of $s_2$ on $\sigma$ for $p=2.5$ and
  $\sigma_0=\sigma$. For $\sigma$ below a given threshold $\sigma_\mathrm{r}$, $s_2$
  becomes negative, indicating that the homogeneous state of motion is
  stable with respect to long wavelength perturbations. In contrast,
  close to the transition line, this state is unstable since
  $s_2>0$. The vertical dashed line corresponds to the transition
  value $\sigma_\mathrm{t}$. Inset: comparison, close to $\sigma_\mathrm{t}$,
  of $s_2$ obtained numerically from the Boltzmann equation (full line)
  and analytically from the hydrodynamic equations (dashed line),
  showing a good agreement.
  (b) Phase diagram indicating, for $\sigma=\sigma_0$,
  the three different regions: no motion
  ({\bf A}), unstable homogeneous motion ({\bf B}), stable homogeneous
  motion ({\bf C}). The full line has been obtained numerically from
  a stability analysis of the Boltzmann equation.
  The dashed one is the transition line shown in Fig.~\ref{phase-diag}(a).
}
\label{fig-s2}
\end{figure}

The integral equations (\ref{eq-G0}) and (\ref{eq-G1}) can be solved
numerically, once expressed in terms of Fourier coefficients.
To this purpose, we use the following Fourier expansion
of the integral operators $I_\mathrm{dif}$ and $I_\mathrm{col}$:
\bea
\int_{-\pi}^{\pi} d\theta \, e^{ik\theta} I_\mathrm{dif}[g] &=&
-\lambda\left(1-e^{-k^2 \sigma_0^2/2}\right) \hat{g}_k \; ,\\
\int_{-\pi}^{\pi} d\theta \, e^{ik\theta} I_\mathrm{col}[g,h] &=&
\frac{2d_0 v_0}{\pi} \sum_{q=-\infty}^{\infty} \left( e^{-k^2\sigma^2/2}
I_{q-\frac{k}{2}}-I_q \right) \hat{g}_{k-q} \hat{h}_q .
\eea
Numerical results are reported in Fig.~\ref{fig-s2}(a), where $s_2$ is
shown as a function of $\sigma$ for $\sigma=\sigma_0$, all other
parameters being kept fixed.  Consistently with the results obtained
from the hydrodynamic equations, we observe that close to the
transition line, $s_2$ is positive and diverging.  But for smaller
values of the noise amplitude $\sigma$, $s_2$ becomes negative. Hence
the homogeneous state of motion becomes stable in this range with
respect to long wavelength perturbations.

As summarized on Fig.~\ref{fig-s2}(b), there are from the point of
view of stability three regions in the phase diagram
(we focus here on the case $\sigma=\sigma_0$).
These three regions can be described as follows:
\begin{description}
  \item[A] At low $p$ or high $\sigma$, no collective motion occurs.
  \item[B] For $p_\mathrm{t}<p<p_\mathrm{r}$, a homogeneous stationary solution with
    nonzero velocity exits, but it is unstable under longitudinal
    compression modes.
  \item[C] For $p>p_\mathrm{r}$, the homogeneous and stationary moving phase is
    linearly stable under any small perturbation.
\end{description}
$p_\mathrm{r}$ is defined as the value of the reduced density such that
$s_2=0$. Note that $p_\mathrm{r}$ is not a monotonous function of $\sigma$.
In the {\bf B} region, the system cannot converge to a homogeneous
stationary solution, and one thus expects the system to organize into
more complicated spatio-temporal structures, that we shall try to
describe in Section~\ref{sec-soliton}.

\subsection{Restabilization of the homogeneous flow in the hydrodynamic equations} \label{sec-restab-hydro}

The above stability analysis from the Boltzmann equation shows that
the homogeneous flow becomes linearly stable when getting farther from
the transition line $p_\mathrm{t}$.
Although this region of restabilization is, strictly speaking, out of the
validity domain of the hydrodynamic equations, it would be interesting
to know whether these equations already contain, at a qualitative level
of description, the restabilization phenomenon.

\begin{figure}
\begin{center}
\psfrag{Re[s]}{$\Re[s]\lambda^{-1}$}
\includegraphics[width=0.75\textwidth,clip]{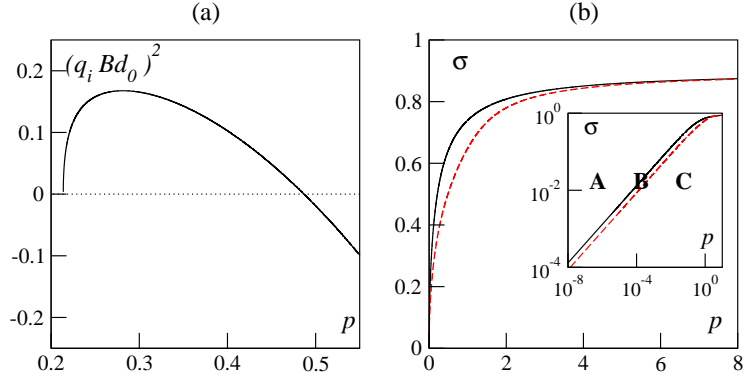}
\caption{Restabilization in the hydrodynamic framework. (a)
  $\left(q_iBd_0\right)^2$ such that $\Re [s]=0$ \emph{versus} $p$,
  same parameters as Fig.~\ref{Fdispersion1}. (b) Phase diagram. The
  full line corresponds to the onset of motion, $\sigma_\mathrm{t}$. The dashed
  line is the transition between stable and unstable homogeneous
  flows, $\sigma_\mathrm{r}$. Regions {\bf A}: $w=0$, {\bf B}: $w\ne 0$ and $\Re
  [s]>0$ when $q < q_i$, {\bf C}: $w\ne 0$ and $\Re [s]<0$ for all
  $\mathbf{q}$ and for all direction of perturbation. Inset: same as
  (b) in log-log scale.}
\label{Fdispersion2}
\end{center}
\end{figure}

One possible way to investigate this stability issue is study the
sign of the coefficient $s_2$ of the $q^2$ term in the small $q$ expansion
of $\Re[s_+]$. An equivalent procedure, that we follow here, is to look
for the domain of existence of the wavenumber $q_i$
(defined as $\Re[s_+]=0$ for $q_i \ne 0$),
when the control parameter $p$ is increased at a given noise amplitude
$\sigma$. In order to achieve this task, we solve the equation $\Re[s_+]=0$,
using expression~(\ref{Eq_Res}). The solutions are naturally expressed
in terms of the variable $q_i^2$. After some algebra, we find for the largest
solution: 
\begin{eqnarray}
\label{eq-qi2}
  q_i^2&=&\frac{\mu}{\nu v_0^2}
    \left[-\gamma w_1^2\left(\frac{\mu'}{\mu}- \frac{\xi'}
      {\xi}\right)-2 v_0^2\right.\\ 
 && \qquad \qquad +\left.w_1\left(\frac{\mu'}{\mu}-\frac{\xi'}{\xi}\right)
      \sqrt{\gamma ^2w_1^2+2 v_0^2} \right],
\nonumber
\end{eqnarray}
where the term $\left(\frac{\mu'}{\mu}-\frac{\xi'}{\xi}\right)$
is positive.
The expression in the right hand side of Eq.~(\ref{eq-qi2}) is positive
for $p$ close enough to $p_\mathrm{t}$, and becomes negative for larger $p$
(see Fig.~\ref{Fdispersion2}(a)),
in which case a real solution $q_i$ does not exists.
As a result, there exists a value $p_\mathrm{r}$ of the control
parameter $p$ such that $q_i$ is no longer defined. For $p>p_\mathrm{r}$,
$\Re[s_+]$ remains negative for all values
of $q$ (Fig.~\ref{Fdispersion1}(a)), so that all perturbations
are linearly stable.
Using equations~(\ref{eq-nu})-(\ref{eq-xi}), we can compute
the restabilization line $p_\mathrm{r}(\sigma)$ and show
that $p_\mathrm{r}$ depends only on $\sigma$ and $\sigma_0$, but not
on $B$. We also find that $p_\mathrm{r}(\sigma)$
behaves for small noise amplitude as $p_\mathrm{r}\propto \sigma^{1/2}$
(see Fig.~\ref{Fdispersion2}(b) and its inset).

Altogether, the hydrodynamic equations seem to lead to the correct
phenomenology even when used beyond their strict validity domain.
Yet, the locations of the transition line $p_\mathrm{r}(\sigma)$ predicted from
the hydrodynamic equations on one side, and the one predicted
from a long wavelength perturbative treatment of the Boltzmann equation
on the other side are quantitatively different, as illustrated on
Figs.~\ref{fig-s2}(b) and~\ref{Fdispersion2}(b).

%%%%%%%%%%%
\subsection{Inhomogeneous travelling solutions and solitary waves}
\label{sec-soliton}

For $p$ slightly larger than $p_\mathrm{t}$, the homogeneous solutions $\mathbf{w}=0$
and $\mathbf{w}=\mathbf{w}_1$ are unstable, and one should look
for the onset of spatio-temporal structures rather than purely
stationary states.  In this respect, one may be guided by the
observations made in numerical simulations
\cite{Gregoire2004,Chate2008}, where 'stripes' of higher density
moving over a low density background have been reported. Such
structures are rather similar to soliton solutions that have been
observed in many different physical contexts \cite{Dauxois}.

\subsubsection{Stationary hydrodynamic equation in a moving frame.}

Let us now look for possible soliton solutions of the hydrodynamic equations
(\ref{continuity}) and (\ref{eq-w}). To this aim, we assume
for $\rho$ and $\mathbf{w}$
the following ``propagative'' form, with propagation velocity $c>0$,
along an arbitrary axis $x$ of unit vector $\mathbf{e}$:
\be
\rho(\mathbf{r},t)=R(x-ct), \quad
\mathbf{w}(\mathbf{r},t)=W(x-ct)\,\mathbf{e},
\ee
with $\zeta=x-ct$ and $W(\zeta)>0$.
Using Eqs.~(\ref{continuity}), one finds the simple relation $R'=W'/c$,
leading to:
\be
R(\zeta) = \frac{1}{c} W(\zeta) + \rho^*,
\ee
where $\rho^*$ is up to now an arbitrary constant density.  In the
following, we consider velocity profiles that vanish for $\zeta \to
\pm \infty$, so that $\rho^* = \lim_{\zeta \to \pm \infty} R(\zeta)$.
Inserting this form in Eq.~(\ref{eq-w}), one can eliminate $R(\zeta)$
and obtain the following ordinary differential equation for
$W(\zeta)$, also taking into account the density dependence of the
transport coefficients
\footnote{We however neglect the density dependence of the ratio $\nu'/\nu$,
as it would lead to terms of higher order than that retained
in our hydrodynamic description.}:
\be \label{eq-soliton}
W''=-(a_0-a_1 W -a_2 W')W' -b_1 W -b_2 W^2 - b_3 W^3.
\ee
The different coefficients in Eq.~(\ref{eq-soliton}) read
\bea
a_0 &=& \left(c-\frac{v_0^2}{2c}\right) (D_1+D_2\rho^*)\\
a_1 &=& \tilde{\gamma} + D_2 \left(\frac{v_0^2}{2c^2}-1\right)\\
a_2 &=& \frac{D_2}{c(D_1+D_2\rho^*)}\\
b_1 &=& \mu' (\rho^*-\rho_\mathrm{t}) (D_1+D_2\rho^*)\\
b_2 &=& \frac{\mu'}{c} \left[ D_1 + D_2 (2\rho^*-\rho_\mathrm{t}) \right]\\
b_3 &=& \frac{\mu' D_2}{c^2} -\tilde{\xi}
\eea
with
\bea
D_1 &=& \frac{4 \lambda}{v_0^2}\left( 1-e^{-2\sigma_0^2} \right)\\
D_2 &=& \frac{64 d_0}{3\pi v_0} \left(\frac{7}{5}+ e^{-2\sigma^2}\right),
\eea
and $\tilde{\gamma}=\gamma/\nu$, $\tilde{\xi}=\xi/\nu$.
As often in the study of solitons \cite{Dauxois}, Eq.~(\ref{eq-soliton})
may be reinterpreted as the equation of motion of a fictive particle
with position $W$ at time $\zeta$. Here, this virtual particle has
a unit mass, and moves in a potential
\be
\Phi(W) = \frac{b_1}{2} W^2 + \frac{b_2}{3} W^3 + \frac{b_3}{4} W^4,
\ee
with a non-linear friction force $-(a_0-a_1 W-a_2 W')W'$.
Depending on the sign of the effective friction coefficient
$(a_0-a_1 W-a_2 W')$, the 
friction force may either dissipate or supply energy to the particle.
Note that this friction term breaks the symmetry $\zeta \to -\zeta$,
so that the resulting momentum profile cannot be symmetric.

\subsubsection{Numerical integration of the velocity and density profiles.}

To find a solution for $W(\zeta)$, we integrate numerically
Eq.~(\ref{eq-soliton}) for given values of the parameters $a_i$ and $b_i$.
The following constraints are imposed to the solution: $W(\zeta)$ should be
positive for all values of $\zeta$, and $W(\zeta)$ should go to $0$
for $\zeta \to \pm \infty$. Hence for large values of $|\zeta|$, $W(\zeta)$
should be small, and should satisfy, to a good accuracy,
the linearized version of Eq.~(\ref{eq-soliton}), namely:
\be \label{soliton-lin}
W''+a_0 W' +b_1 W=0.
\ee
This equation has two exponential solutions $W_{\pm}(\zeta)=
A_{\pm}\exp(r_{\pm}\zeta)$, with:
\be
r_{\pm} = \frac{1}{2}\left(-a_0 \pm \sqrt{a_0^2-4b_1}\right).
\ee
For $W(\zeta)$ to be positive, one needs that the roots $r_{\pm}$ be real,
which implies $a_0^2-4b_1 \ge 0$. Further, for $W(\zeta)$ to vanish both
for $\zeta \to -\infty$ and $\zeta \to +\infty$, one should have both
an increasing and a decreasing solution for Eq.~(\ref{soliton-lin}), namely
$r_{+}>0$ and $r_{-}<0$, which corresponds to $b_1<0$.

The free parameters in Eq.~(\ref{eq-soliton}) are $c$ and $\rho^*$
(this point will be briefly discussed in Section~\ref{solitons_num},
in connection with numerical results). The noises
$\sigma$ and $\sigma_0$ are external control parameters. The overall density 
$\rho$ is computed afterwards from the profile $R(\zeta)$.
Assuming that we are in the low noise region of parameter space
$\sigma<\sigma_\mathrm{t}^{\infty}$, then $[\exp(-\sigma^2/2)-2/3]>0$ and
the condition $b_1<0$ implies $\rho^*<\rho_\mathrm{t}$.
In addition, as the trajectory of the particle starts and ends at the same
position $W=0$ with zero velocity ($W'=0$),
its energy is the same, which means that
the friction force has to dissipate energy on some part of the trajectory
and to supply energy otherwise.
Assuming $a_0>0$ implies $c-v_0^2/(2c)>0$, that is $c>v_0/\sqrt{2}$.
On the other hand, one intuitively expects $c$ to be smaller than the
microscopic velocity $v_0$ of the particles.

\begin{figure}[t]
\centering\includegraphics[width=0.75\textwidth,clip]{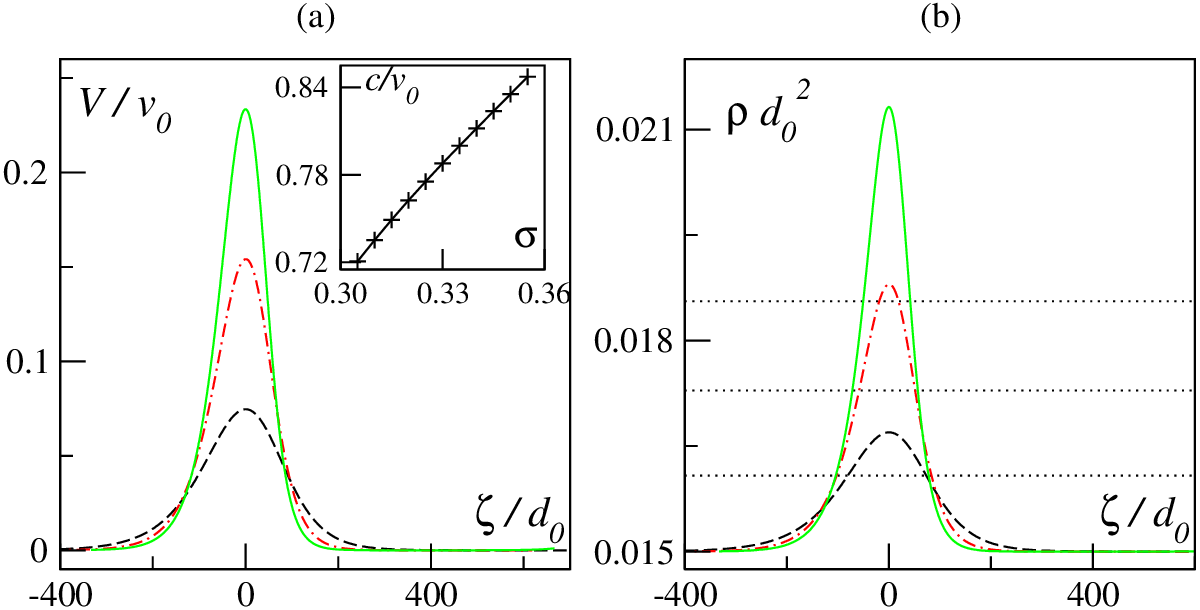}
\caption{(a) Velocity profile $v(x,t)=V(\zeta)$,
with $\zeta=x-ct$, for $\rho^*=0.06$ and
$\sigma=\sigma_0=0.31$ (dashed line), $0.32$ (dot-dashed line)
and $0.33$ (full line). Inset: propagation velocity $c$ as
a function of $\sigma$.
(b) Density profile $\rho(x,t)=R(\zeta)$ for the
same values of the parameters. Horizontal dotted lines correspond
to the density $\rho_\mathrm{t}$ for $\sigma=\sigma_0=0.31$, $0.32$ and $0.33$
(bottom to top). Other parameters: $\lambda=0.5$, $d_0=0.5$ and $v_0=1$.
}
\label{fig-soliton}
\end{figure}

The numerical procedure we implement is the following.
Choosing a given value for $\rho^*$ and for $c$,
we start at $\zeta=\zeta_0<0$ ($|\zeta_0| \gg 1$),
with a small value $W(\zeta_0)=W_0 \ll 1$,
and with a derivative $W'(\zeta_0)=r_{+}W_0$.
This choice of initial conditions ensures that we select a solution
with an exponential tail
$W(\zeta) = A_{+}\exp(r_{+}\zeta)$ for $\zeta < \zeta_0$.
Then we integrate numerically the differential equation for $\zeta>\zeta_0$,
until reaching large enough positive values of $\zeta$.
At this stage, two behaviours may appear. One should first notice
that for $b_1<0$ (and at least if $b_3$ does not take a large negative value)
the effective potential $\Phi(W)$ has a local maximum in $W=0$
and a local minimum at a value $W=W_{\rm min}$.
Then, if the dissipated energy is larger than the injected energy, the
particle ends up at $W_{\rm min}$ for $\zeta \to \infty$, yielding
a profile $W(\zeta)$ that does not fulfil the condition required.
In the opposite case, if energy injection dominates, the particle crosses
the local maximum at $W=0$ and goes to negative values.
It is only in the marginal case where dissipation exactly compensates
injection that the correct profile $W(\zeta)$ is found.
As friction is mainly controled by the parameter $c$, we keep $\rho^*$
fixed and perform a loop over the value of $c$ in order to converge toward
the marginal solution. Note however that if $b_3<0$, $\Phi(W) \to
-\infty$ when $W \to +\infty$, so that one should also take care
that the particle does not ``escape'' to large positive values of $W$.

Using the above procedure, we obtain a family of profiles $W(\zeta)$
with three control parameters, namely the ``background'' density $\rho^*$
and the noises $\sigma$ and $\sigma_0$. In the following, we restrict
ourselves to the case $\sigma_0=\sigma$.
The density profile is computed from the relation
$R(\zeta)=\rho^*+W(\zeta)/c$, and the velocity profile $V(\zeta)$ is
obtained from the momentum profile $W(\zeta)$ through
$V(\zeta)=W(\zeta)/R(\zeta)$. Examples of such velocity profiles
are presented in Fig.~\ref{fig-soliton}, for different values
of $\sigma$ and for a given value of $\rho^*$.

A remaining open issue is the stability of these solitary waves with respect
to small perturbations. A formal stability analysis like the one performed
for the homogeneous state of motion is a difficult task here, and we leave
this question for future work.

%%%%%%%%%%%
\subsection{Solitary waves in the agent-based model}
\label{solitons_num}

\begin{figure}[t]
\centering\includegraphics[width=0.75\textwidth,clip]{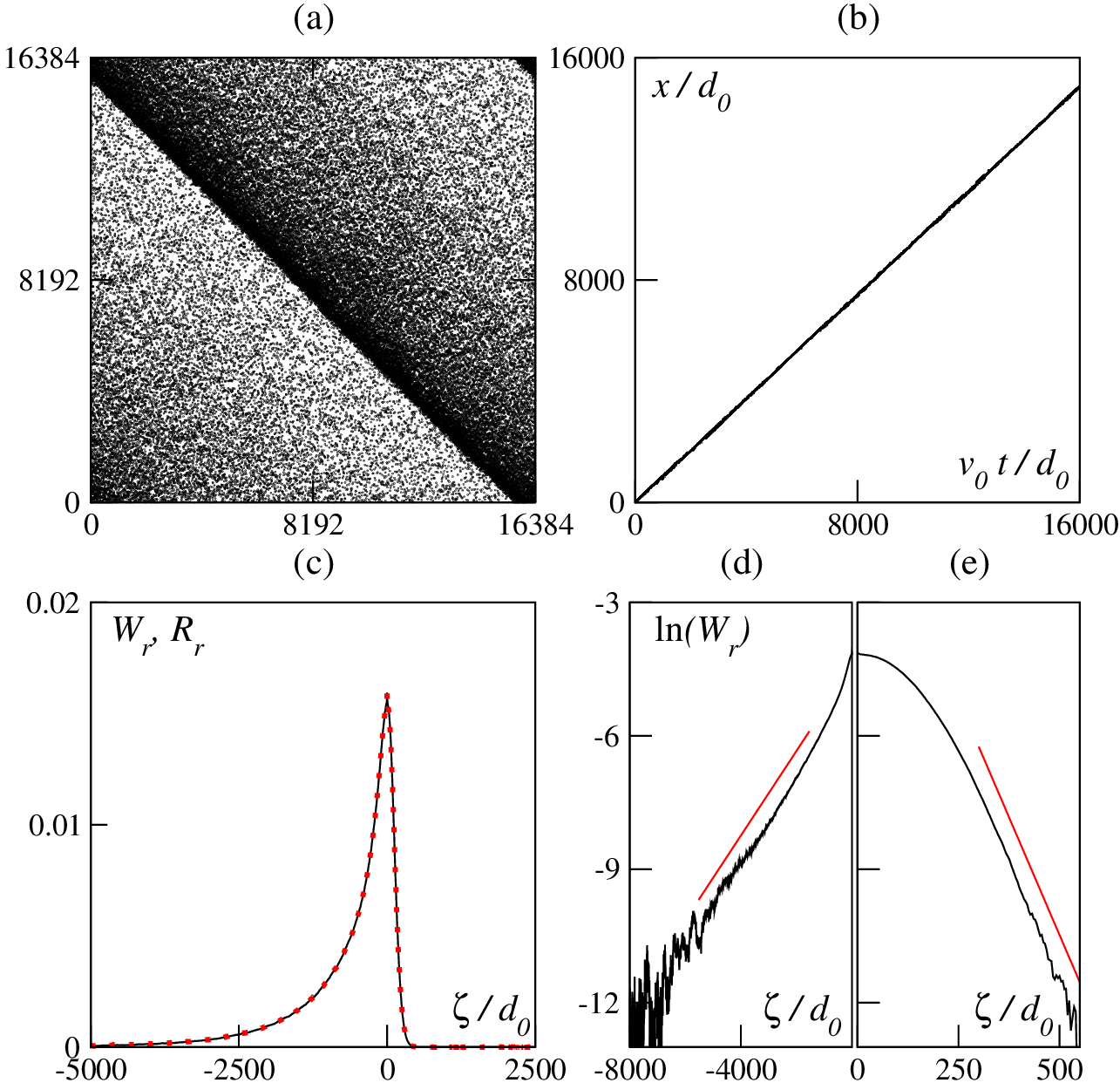}
\caption{Solitons in the numerical model. (a) Instantaneous snapshot, the
  band is moving south-west; lengths are scaled by $d_0$.
  (b) Example of trajectory in the
  direction of the averaged velocity. (c) Mean profiles along the
  direction of the main motion. We plot the reduced dimensionless
  density
  $R_\mathrm{r}=(\langle\rho(x-ct)\rangle-\rho^\mathrm{sat})d_0^2 c/v_0$ 
  (dotted line) and the dimensionless momentum $W_\mathrm{r}=\langle
  w(x-ct)\rangle d_0^2/v_0$ (plain line), both being time-averaged
  in the comoving frame of the soliton.
  (d) and (e) Same data as (c) on semi-log scales,
  emphasizing the exponential decay.
  The scales are identical on vertical axes, but different on abscissas.
  Parameter values are $p=2^{-3}$, $\sigma=0.163$, $L=4096$;
  the other ones correspond to set VII in Table~\ref{Tparam}.}
\label{sol_num1}
\end{figure}

\begin{figure}[t]
\centering\includegraphics[width=0.75\textwidth,clip]{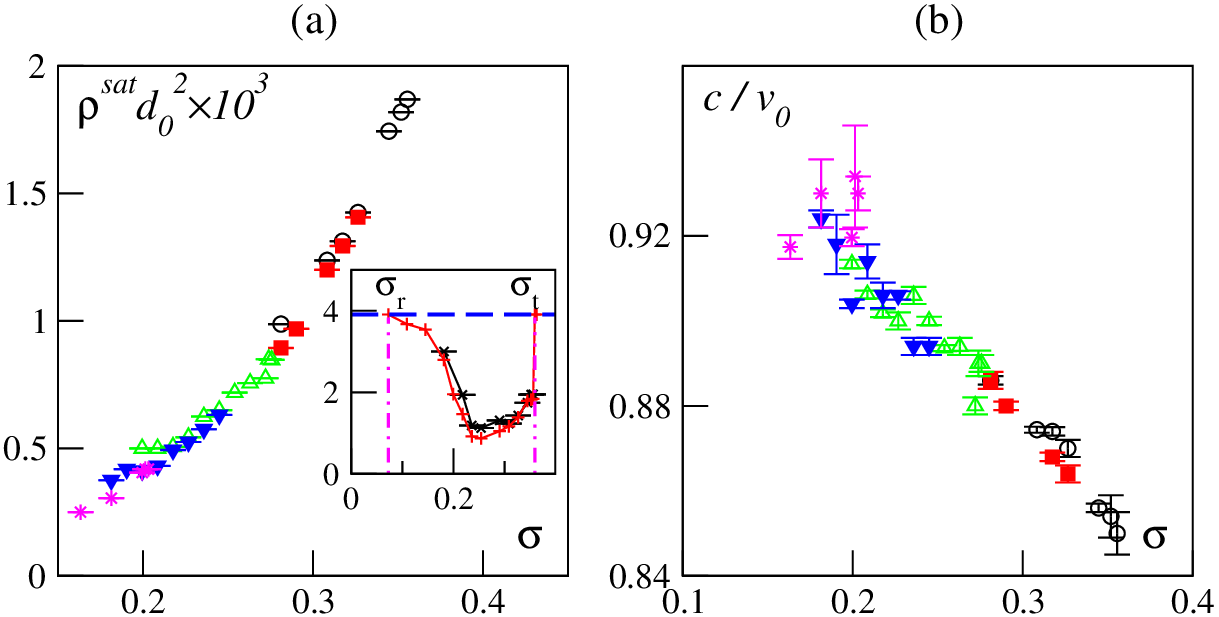}
\caption{Solitons in numerical model. (a) Density of saturated vapour
  for different $p$ ($p=2^{-2},\,\circ$, $5^{-1},\,\blacksquare$,
  $2^{-3},\,\vartriangle$, $10^{-1},\,\blacktriangledown$ and
  $2^{-4},\,*$). Inset: finite size effects on soliton, $p=2^{-4}$,
  $L=1024$ $\times$ and $L=2048$ $+$. The dashed line marks the value
  of the global density ($\rho=2^{-5}$). The dotted lines underline
  the threshold of the collective motion $\sigma_\mathrm{t}$ and of the
  homogeneous moving population $\sigma_\mathrm{r}$.(d) Speed of
  the solitons (same parameters). The other parameters are the ones of
  set VII (see table~\ref{Tparam}).}
\label{sol_num2}
\end{figure}

We now compare the solitary waves computed in the hydrodynamic
equations with the travelling stripes observed in direct numerical
simulations of the agent-based model (see Fig.~\ref{sol_num1}(a)). We
focus again on the case $\sigma_0=\sigma$.  The stripped structures
are composed by several localised, randomly spaced bands. They are not
part of a regular pattern, nor a wave train~\cite{Chate2008}. They are
all moving along the direction of the main motion, although during the
transient period they can pass through each other with only few
interactions.  The space between two bands is filled with particles
moving independently (the hydrodynamic momentum vanishes), and
homogeneously (the density is constant). In analogy to the liquid-gas
coexistence, we denote this state as the \emph{saturating vapour}.

We observe that the bands move at a constant speed, at least on the
duration necessary to travel through the system size
(Fig.~\ref{sol_num1}(b)). From the trajectories, we measured the
velocity $c$ of the solitons. On the density profiles, we extracted
the value $\rho^{\rm sat}$ of the density outside the peak.  If these
structures are only propagative and if the continuity equation is
valid at a coarse-grained level in the agent based model
(which is expected from mass conservation), the density
and momentum profiles should be related by $W=c(R-\rho^{\rm sat})$, as in
Section~\ref{sec-soliton}. Plotting on Figure~\ref{sol_num1}(c) both
the reduced density $c(R-\rho^{\rm sat})$ and the momentum $W$, we
observe that both curves match perfectly, confirming the propagative
nature of this stripped pattern.

These solitary waves are quite similar to the soliton we found in the
hydrodynamic equations (see section~\ref{sec-soliton}),
with in particular an exponential decay of the momentum profile
on both sides (Fig.~\ref{sol_num1}(d-e)).
However, the asymmetry of the profile is much more pronounced than
in the analytical model: the exponential decay is much steeper
in front of the profile than in the rear part.

We now study how the two main characteristics of the solitary waves,
namely the velocity $c$ and the density $\rho^{\rm sat}$, vary with
the control parameters $p$ and $\sigma$.  Since we perform a numerical
study, we need to be aware of finite size effects. Plotting the
density of saturating vapour versus noise for a given density but for
different sizes, we can make three observations (inset of
Fig.~\ref{sol_num2}). First, the system size hardly changes $\rho
^{\rm sat}$ provided that the noise amplitude remains near the
threshold. Moreover, at a lower noise amplitude, the density $\rho
^{\rm sat}$ increases and become sensitive to the system size. Lastly,
there is a noise $\sigma_{\rm r}$ below which we cannot
observe solitons anymore (see also~\cite{Chate2008}) and the system
becomes homogeneous at a coarse-grained level. The result is
qualitatively consistent with the restabilization of the homogeneous
flow described in Sections~\ref{sec-stab-boltz} and
\ref{sec-restab-hydro}.

The study of the very low noise amplitude region of the phase diagram
is an ongoing work. So we mainly focused in the present article on the
region relatively close to the transition to collective motion.  For
different global densities, both $\rho^\mathrm{sat}$ and $c$ fall onto
the same curve when plotted as a function of $\sigma$, as shown on
Fig.~\ref{sol_num2}(a) and (b).  Therefore, once the noise amplitude
$\sigma$ is given, the characteristics $(c,\rho^\mathrm{sat})$ of the
solitary waves are determined, and the number of solitary waves is
adjusted by the dynamics in order to match the global density of the
system.

This is a major difference with the solitary waves obtained from the
hydrodynamic equations in section~\ref{sec-soliton}.
These solitary waves depend on two control parameters, namely the noise
amplitude $\sigma$ and the density at infinity $\rho^*$.
Hence there is a priori no way to determine the number of solitons
in a large but finite system with a given density.
At a heuristic level, we might guess that the solitary waves may be stable
only for some specific values of $\rho^*$, which would give
a selection mechanism for the density $\rho^\mathrm{sat}$.
Such a mechanism would make the connection between the analytical
and numerical models clearer, but we presently have no clue to confirm
this tentative scenario.
Obviously, further studies of the dynamics of the solitary waves
in the context of the hydrodynamic equations are needed.

%%%%%%%%%%%%%%%%%%%%%%%%%%%%%%%%%%%%%%%%%%%%%%%%%%%%%%%%%%%%%%%%

\section{Conclusion}
\label{sec-discussion}

In summary, we have derived in this article hydrodynamic equations for a model
of self-propelled particles with binary interactions, in the
regime of low hydrodynamic velocity.
We also compared the results of the hydrodynamic description to the
numerical simulations of a standard agent-based model.
In the analytical model,
the homogeneous state with zero velocity is a stationary solution
for any values of the microscopic parameters (the noise amplitude
and the overall density), but this state is linearly
unstable for a reduced density $p$ greater than a transition density
$p_\mathrm{t}(\sigma)$, or equivalently, for a noise smaller than a transition
value $\sigma_\mathrm{t}(p)$.

When the zero velocity solution is unstable, another homogeneous state,
with a nonzero hydrodynamic velocity, appears.
This state is linearly stable with respect to spatially homogeneous
perturbations.
However, close to the transition line $\sigma_\mathrm{t}(p)$, this state turns
out to be linearly unstable with respect to finite wavelength perturbations.
As the validity of the hydrodynamic equations is, strictly speaking,
restricted to the vicinity of the transition line, we also studied the
stability of the homogeneous state of motion directly from the Boltzmann
equation. We found that, far enough from the transition line, the
homogeneous motion becomes linearly stable.
Interestingly, this restabilization phenomenon is also qualitatively
observed in the hydrodynamic equations, although this regime is
beyond their domain of validity.
All these results agree semi-quantitatively with the numerical
simulations of the agent based model.

When the homogeneous state of motion is unstable, more complex
spatio-temporal structures should appear. A candidate for such structure
is the solitary waves we obtained from the hydrodynamic equations.
These solitary waves resemble the moving stripes observed in
the numerical agent-based model, apart from the asymmetry
which is more pronounced in the latter.
A word of caution is however needed here, as on the one hand the
solitary waves have a finite amplitude, so that the hydrodynamic
equations might not be valid, and more importantly, their stability
has not been tested yet. On the basis of the numerical simulations
of the agent-based model, it is however likely that these solitary waves
should be stable at least in a given region of the phase diagram.

As for future work, it would be interesting to investigate
the stability of the solitary waves, and to look for possible
``multi-soliton'' solutions, in case the stability would be confirmed.
Specifically, it would be interesting to be able to determine the
number of solitons, their celerity and the background density
as a function of the global density (for a finite volume) and
of the noise amplitude, if such a relation exists, as suggested
by the numerical simulations of the agent-based model.

\section*{Acknowledgements}

This research work has been partly supported by the
French ANR project ``DyCoAct'' and by the Swiss National Science
Foundation.

%%%%%%%%%%%%%%%%%%%%%%%%%%%%%%%%%%%%%%%%%%%%%%%%%%%%%%%%%%%%%%%%
%%%%%%%%%%%%%%%%%%%%%%%%%%%%%%%%%%%%%%%%%%%%%%%%%%%%%%%%%%%%%%%%
%%%%%%%%%%%%%%%%%%%%%%%%%%%%%%%%%%%%%%%%%%%%%%%%%%%%%%%%%%%%%%%%

\appendix{}

%%%%%%%%%%%%%%%%%%%%%%%%%%%%%%%%%%%%%%%%%%%%%%%%%%%%%%%%%%%%%%%%

\section{Agent-based model}
\label{app-num-model}

%%%%%%%%%%%
\subsection{Looking at the model further}

The numerical system we looked at is very similar to the one defined
by Vicsek \emph{et al}~\cite{Vicsek1995}. This is a very minimal
model, easy to implement. In contrast, a real direct simulation would
have been coded following a molecular dynamics algorithm, which would have
cost much more cpu time than our Monte Carlo-like program. The
numerical choice is also related to the fact that collective motion of
self-propelled particles has been mainly studied in this framework
during the last ten
years~\cite{Albano,Aldana2007,Huepe,Chate2008,Couzin2002,Czirok1997,Duparcmeur,Gregoire2001,Gregoire2004,Levine,Mikhailov,Ken}. Thus
we would like to take profit from this large background and the
knowledge of the system we already got.

To fully understand the results presented in this paper, we must explain the
differences between the numerical system we used and a direct
simulation. In what we have done, collisions are computed at fixed
time step. So every other collision that could have occurred within
$\Delta t$ is neglected. On the other hand, collisions can involved
many individuals. Another implication of the discrete time step is
that decreasing the time step increases the collision frequency. Then
the noise does not act on the system with the same manner for two
different time steps. Hence, in its present formulation, the agent-based
model is not a discretized version of a continuous time model.
To reach this goal,
the noise amplitude should be renormalized in some way with the time step.

The balance of the above different effects is difficult to imagine a
  priori. We do not expect any \emph{quantitative} matching between the
theory we developed and the simulations we
presented. But we still want to test the robustness of the predictions
made for large system sizes.

We must also emphasize that some studies
in the literature were aimed at
giving an exact continuous theory of Vicsek's
model~\cite{Ratushnaya,Degond,Degond2008}. Up to now, this difficult
problem has been dealt in the framework of perturbative theories at
a first order in speed differences. 
In addition, the role of the noise is not properly taken into account
in these studies: it is either ignored \cite{Ratushnaya},
or described by a phenomenological diffusive term~\cite{Degond}. Finally, the
transport coefficients of the hydrodynamic equation do not contain
any dependence on the microscopic parameters of the model.

\begin{figure}[t]
\begin{center}
\includegraphics[width=0.75\textwidth,clip]{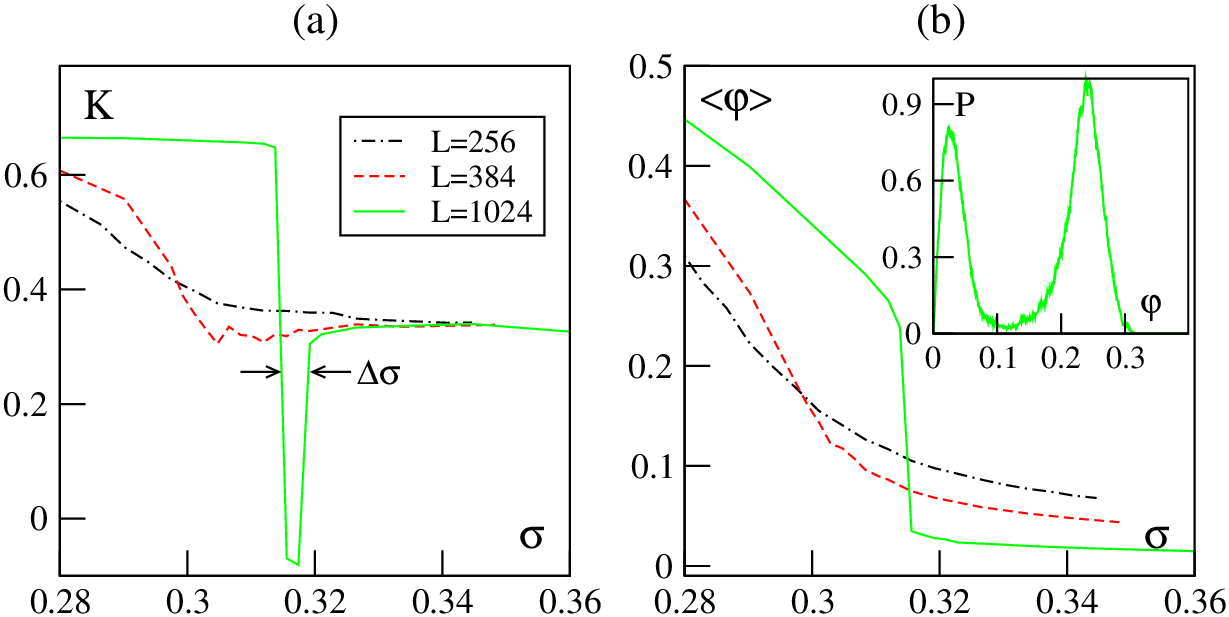}
\caption{Phase transition in numerical simulation and finite size
  effects. We plotted (a)~the Binder cumulant $K$ and (b)~the averaged
  order parameter $\langle\varphi\rangle$ \emph{vs} noise
  \emph{rms}-value, for three different sizes. In the inset, we show
  the histogram of the order parameter $\varphi$ at the transition
  point for a system size $L=1024$. On figure (a), we emphasized the
  depth of the well $\Delta \sigma$: approximation of the errors in
  determining the transition point. Parameters are the ones of
  configuration n$^\circ$ III with $\rho=1/16$ or $p=1/2$.}
\label{tran_num}
\end{center}
\end{figure}

%%%%%%%%%%%
\subsection{Phase transition}

As in usual versions of self-propelled particles systems, the
behaviour of the system roughly falls into two different categories.
Either there is no collective motion: every particle move randomly without
clear correlation with its neighbours; or there is a non-zero global
velocity in an arbitrary direction.

Since the analogy with magnetic systems is quite obvious, the habits
is to consider the equivalent averaged magnetization of our system,
namely the global normalized velocity $\varphi^t$:
\be
\varphi^t=\left|\left| \frac{1}{Nv_0}\sum_{j=1}^N\mathbf{v}_j^t\right|\right|,
\ee
considered as an order parameter.
To determine the characteristics of the phase transition, we study the
statistical properties of the order parameter $\varphi^t$,
considering its mean $\langle\varphi\rangle$, its variance $\chi$
and its Binder cumulant $K$~\cite{Binder}:
\bea
\langle\varphi\rangle &=& \frac{1}{T}\sum_{t=1}^T\varphi^t,\\
\chi &=& L^2\left(\langle\varphi^2\rangle-\langle\varphi\rangle^2\right),\\
K &=& 1-\frac{\langle\varphi^4\rangle}{3\langle\varphi^2\rangle^2}.
\eea
The brackets $\langle \ldots \rangle$ indicate an averaging over time.
The duration of the simulation has to be large to inhibit memory effects.
Ideally, the correlation time for each set of parameters $(\rho, \sigma,
\sigma_0, v_0, d_0, \lambda)$ should be computed from the
auto-correlation function~\cite{Muller}. However, this is a tantamount
task\footnote{The cumulative consumed cpu time already reaches fifty
years.}.
Practically, in order to have a rough approximation of the correlation
time, we measured the transition time from the initial condition to
the stationary state. Then we performed averaging on time which are
hundred times greater than that transition time.

For all sets of parameters I to VII (Table~\ref{Tparam}), we observed
that the system exhibits a phase transition from a non-moving to a
globally moving population when decreasing the noise
amplitude at a fixed density. At small enough size $L$, all
statistical variables $(\langle\varphi\rangle, \chi, K)$ remain
continuous, while a singular point appears when the system is
larger than a typical size $L_\mathrm{t}$ (see Fig.~\ref{tran_num}(a) and (b),
as well as Refs.~\cite{Gregoire2004,Chate2008}).

The main observations are the following:
the order parameter curve exhibits a jump
(Fig.~\ref{tran_num}(b)), the variance is delta-peaked (not shown
here), the Binder cumulant has a minimum (Fig.~\ref{tran_num}(a))
which goes to larger negative values when the system size is increased,
and the histogram of the order parameter is bimodal
(see inset of figure~\ref{tran_num}(b)).
All these sign plead in favour of a first-order phase transition.

It is now well known that a finite size system exhibits a rounded
transition, at equilibrium~\cite{Privman,Borgs} or far from the
equilibrium~\cite{Lubeck,Marcq}. The scaling laws are normally
sufficient to detect the order of the transition. In our case, the
finite size scaling laws correspond to a continuous transition below
$L_\mathrm{t}$~\cite{These_gg,Baglietto}, and to a discontinuous
transition above $L_\mathrm{t}$~\cite{Chate2008}.

To estimate the transition point, we measured the
location where the Binder cumulant minimum becomes negative. We
neglected the finite size effects at higher size. We determined the error
bars on that location as the width of the well (see
Fig.~\ref{tran_num}(a)).

%%%%%%%%%%%%%%%%%%%%%%%%%%%%%%%%%%%%%%%%%%%%%%%%%%%%%%%%%%%%%%%%%%%%%%%%

\section{Stability against arbitrary perturbations}
\label{app-stability}

In this appendix, we study within the framework of the hydrodynamic
equations the stability of the stationary homogeneous flow.
Starting from Eqs.~(\ref{lin-rho}) and (\ref{lin-w}),
we consider the case where
$\mathbf{w}_0=\mathbf{w}_1\ne\mathbf{0}$, solution of
Eq.~(\ref{eq-homogstat}). The rotational symmetry is broken
when the collective motion appears, and we take
$\mathbf{e}_{\parallel}=\mathbf{w}_1/|\mathbf{w}_1|$ as a first vector
of the geometrical basis. Then we define two angles $\vartheta_1$ and
$\vartheta_2$ between $\mathbf{e}_{\parallel}$ and the
directions of $\delta\mathbf{w}_0$ and $\mathbf{q}$ respectively.
We denote as $\mathbf{e}_{\perp}$ the unit vector orthogonal to
$\mathbf{e}_{\parallel}$, and such that
$(\mathbf{e}_{\parallel},\mathbf{e}_{\perp})$ form a direct basis.

From Eqs.~(\ref{lin-rho-sq}) and (\ref{lin-w-sq}), we project
the resulting vectorial equation onto $\mathbf{e}_{\parallel}$ and
$\mathbf{e}_{\perp}$,
and we eliminate the ratio $\delta\mathbf{w}_0/\delta\rho_0$
from the continuity equation, yielding:
\begin{eqnarray}
  &&s\left[s+i\gamma w_1q\cos\vartheta_2
    +\nu q^2\right]\cos\vartheta_1 =\label{Eqcompos_x}\\ 
  &&\qquad\qquad \left[-\frac{1}{2} v_0^2
    q\cos(\vartheta _1- \vartheta _2)+is\kappa
    w_1\cos\vartheta_1 \right]q \cos\vartheta_2 \nonumber\\ 
  &&\qquad\qquad -\left[2s\xi w_1^2\cos\vartheta_1+
    i\left(\mu'-\xi 'w_1^2+s\kappa\right)q
    w_1\cos(\vartheta_1-\vartheta_2)\right],\nonumber\\ 
  &&s\left[s+i\gamma w_1q\cos\vartheta_2 +\nu
    q^2\right]\sin\vartheta_1 = \label{Eqcompos_y}\\ 
  &&\qquad\qquad \left[-\frac{1}{2}v_0^2
    q\cos(\vartheta _1 -\vartheta _2)+is\kappa
    w_1\cos\vartheta_1 \right]q\sin\vartheta_2 ,\nonumber
\end{eqnarray}
where $q$ and $w_1$ are real and positive. These are two polynomial
equations that we will study at a given point $(\rho, \sigma)$ of
the phase diagram, for a set of physical variables $(d_0, \lambda,
v_0)$ and for different pairs $(\vartheta_1,\vartheta_2)$. For all
fixed parameters, the solutions of those equations will be a discrete
number of sets $(q,s)$.

First, one can check that this set of equations is invariant when
$\delta\mathbf{w}_0$ is rotated with an angle of $\pi$
($\vartheta_1 \rightarrow \vartheta_1+\pi$). Note also that
every real term depends on an even power of $q$. So one can
expect that the real part of the growth rate $\Re[s]$ only depends on
even powers of $q$, and that $\Re[s]$ remains invariant when $\vartheta_2$
is changed into $\vartheta _2+\pi$. That is why we will study
Eqs.~(\ref{Eqcompos_x}) and (\ref{Eqcompos_y}) for $(\vartheta
_1,\vartheta_2)\in[0,\pi[\times[0,\pi[$.

A third property arises clearly when we introduce the expressions
(\ref{eq-nu})-(\ref{eq-xi})) of the transport coefficients in
Eqs.~(\ref{Eqcompos_x}) and (\ref{Eqcompos_y}):
the wavenumber $q$ appears only through the product $qB_0 d_0$,
meaning that the solutions $q$ are proportional to $1/Bd_0$.
As already mentioned, the
framework of the kinetic approach implies that $B$ is large, and
therefore it implies that we are studying long wavelength
perturbations. This analysis also shows that the growth rate depends
only on the dimensionless control parameter $p$, the noise amplitudes
$\sigma$ and $\sigma_0$, and the self-diffusion rate $\lambda$ which
gives the proper unit to $s$. Let us also mention that a trivial
solution of the system of equations is $s=0$ for $q=0$. This solution
is actually an artefact of the calculation procedure (namely a
multiplication by $s$), as it is not a solution of the original
equations (\ref{lin-rho}) and (\ref{lin-w}).  Hence we will not
consider this extra solution in the following.

For some parameters $(\vartheta_1,\vartheta_2)$, one or several terms can
vanish, and the degree of the polynomials may decrease. We will first
study the general equations and those particular cases will be
considered later. 
If we combine ((\ref{Eqcompos_x})~$\times\sin\vartheta_1
-$~(\ref{Eqcompos_y})$\times\cos\vartheta_1$), we get a linear
equation in $s$~:

\begin{eqnarray}
  s=&-&\frac{q\cos(\vartheta _1-\vartheta _2)}{2w_1}\times\label{Eqs_q}\\
  &&\times\frac{q v_0^2 \sin(\vartheta _1-\vartheta _2)+
    2iw_1\left(\mu'-\xi 'w_1^2\right)\sin\vartheta_1}
  {2\xi w_1\cos\vartheta_1 \sin\vartheta_1 +iq\kappa\sin\vartheta_2 }
\nonumber
\eea

\begin{figure}
\begin{center}
\psfrag{t2}{$\vartheta _2$}
\psfrag{t1}{$\vartheta _1$}
\psfrag{R}{$\Re[s]\lambda ^{-1}$}
\includegraphics[width=0.75\textwidth,clip]{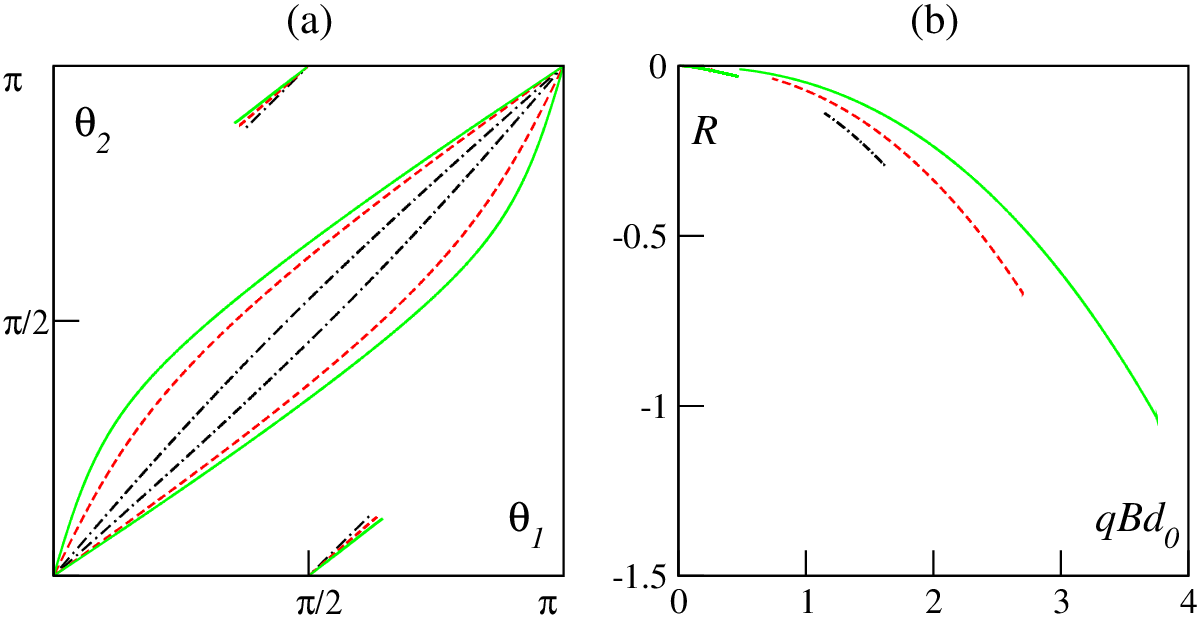}
\caption{Stability against inhomogeneous perturbations, solutions of
  Eqs~(\ref{Eqs_q}) and~(\ref{Eq_qtheta}) so that $\Im [q]=0$. (a) 
$(\vartheta _1, \vartheta _2)$, with $\sigma=\sigma _0=0.5$,
  $p=0.22$ (dot-dashed line), $0.30$ (dashed), $0.40$ (full line).
  (b) $\Re [s]$ \emph{vs} $q$, same parameters as (a), $\Re[s]$
  increases when $p$ increases.}
\label{Fvectoriel}
\end{center}
\end{figure}

Indeed, we verify that $\Re[s]$ is an even function of $q$. Now we can
replace $s$ by its expression in Eq.~(\ref{Eqcompos_y}). The resulting
equation is a third degree polynomial, that can be formally written
as:
\be
 d_3 q^3 + id_2 q^2 + d_1 q + id_0 = 0,
\label{Eq_qtheta}
\ee
where the coefficients $d_i$ are real functions of
$(p,\sigma,\sigma_0,B)$ and of $(\vartheta _1,\vartheta _2)$. The last
three coefficients are rather difficult to manipulate. But this
equation can be easily solved, using Cardano's method for instance. In
the case where
\be
d_3=\sin\vartheta_1 \sin\vartheta_2 \sin(\vartheta_1-\vartheta_2)
\cos(\vartheta _1-\vartheta_2)
\ee
does not vanish, we compute the solutions. The resulting values
$q$ are complex numbers, so that they cannot correspond to physical
solutions. Yet, for some sets of angles $(\vartheta_1,\vartheta _2)$,
the solutions for $q$ are real. We are interested only in
these modes. Determining the angles $(\vartheta _1,\vartheta _2)$
for which $q$ is real, we then compute the growth rate $\Re[s_+]$ using
Eq.~(\ref{Eqs_q}). There are four different branches (see
Figs.~\ref{Fvectoriel}(a)), whose lengths increase when the
control parameter is chosen deeper in the collectively moving
phase (\emph{i.e.} at low $\sigma$ or at high $p$). For all sets of
parameters for which we have computed the growth rate, its real part
remains negative (Fig.~\ref{Fvectoriel}(b)). Thus the homogeneous
moving phase is stable against finite wavelength perturbations in the
general case.

The above calculation relies on the assumption that 
$d_3\ne 0$.
This assumption is not valid in either of the four following cases:
\begin{itemize}

  \item a longitudinal instability ($\sin\vartheta_1 =0$),

  \item a wave vector $\mathbf{q}$ colinear to the direction of the main
    motion ($\sin\vartheta_2 =0$),

  \item a perturbation $\delta\mathbf{w}_0$ colinear to the wave vector 
    $\mathbf{q}$ ($\sin(\vartheta_1-\vartheta_2)=0$).

  \item a perturbation $\delta\mathbf{w}_0$ perpendicular
    to the wavevector $\mathbf{q}$ ($\cos(\vartheta_1-\vartheta_2)=0$).

\end{itemize}

We first consider the study of stability under a longitudinal
perturbation: $\mathbf{w}_1$ and $\delta\mathbf{w}_0$ are
colinear. Then the Eq.~(\ref{Eqcompos_y}) vanishes in two cases:
\be \nonumber
  s=-\frac{iq v_0^2 \cos\vartheta_2}{2\kappa w_1} \; , \quad
  \mathrm{or} \quad \sin\vartheta_2 =0.
\ee
Replacing $s$ by the first expression in
equation~(\ref{Eqcompos_x}), we can show that there is no authorised
mode, in other words $\Im[q]\ne 0$. So, Eq.~(\ref{Eqcompos_y})
vanishes only for $\vartheta _1=\vartheta _2=0$.
The corresponding stability analysis is presented in details in
Section~\ref{stab-hmg}.

For any of the last three cases, we solve equations~\ref{Eqcompos_y}
and~\ref{Eqcompos_y}, and we find that either there is no authorized mode
($q$ is complex), or $\Re[s_+] \le 0$. Thus none of those cases is related to
an unstable mode.

To sum up, this study of the stability of the homogeneous
stationary moving phase shows that the longitudinal direction is the
only mode which can be unstable. This result is consistent with the
observations made in numerical simulations~\cite{Gregoire2004,Chate2008}.

\section*{References}
\bibliography{birds_long_final}

\providecommand{\newblock}{}
\begin{thebibliography}{10}
\expandafter\ifx\csname url\endcsname\relax
  \def\url#1{{\tt #1}}\fi
\expandafter\ifx\csname urlprefix\endcsname\relax\def\urlprefix{URL }\fi
\providecommand{\eprint}[2][]{\url{#2}}

\bibitem{Toner_rev}
Toner J, Tu Y and Ramaswamy S 2005 {\em Annals Of Physics\/} {\bf 318} 170

\bibitem{Parrish1997}
Parrish J~K and Hamner W~M (eds) 1997 {\em Animal Groups in Three Dimensions\/}
  (Cambridge: Cambridge University Press)

\bibitem{Helbing_nat}
Helbing D, Farkas I and Vicsek T 2000 {\em Nature\/} {\bf 407} 487

\bibitem{Helbing}
Helbing D, Farkas I~J and Vicsek T 2000 {\em Phys. Rev. Lett.\/} {\bf 84} 1240

\bibitem{Feare1984}
Feare C 1984 {\em The Starlings\/} (Oxford: Oxford University Press)

\bibitem{Hubbard2004}
Hubbard S, Babak P, Sigurdsson S and Magnusson K 2004 {\em Ecol. Model.\/} {\bf
  174} 359

\bibitem{Rauch1995}
Rauch E, Millonas M and Chialvo D 1995 {\em Phys. Lett. A\/} {\bf 207} 185

\bibitem{BenJacob1995}
Ben-Jacob E, Cohen I, Shochet O, Czir\'ok A and Vicsek T 1995 {\em Phys. Rev.
  Lett.\/} {\bf 75} 2899

\bibitem{Harada1987}
Harada Y, Nogushi A, Kishino A and Yanagida T 1987 {\em Nature (London)\/} {\bf
  326} 805

\bibitem{Badoual2002}
Badoual M, J\"ulicher F and Prost J 2002 {\em Proc. Natl. Acad. Sci. USA\/}
  {\bf 99} 6696

\bibitem{Sugarawa2002}
Sugawara K, Sano M and Watanabe T 2002 {\em Proc. of 2002 FIRA Robot World
  Congress\/}  36

\bibitem{Vicsek1995}
Vicsek T, Czir\'ok A, Ben-Jacob E, Cohen I and Shochet O 1995 {\em Phys. Rev.
  Lett.\/} {\bf 75} 1226

\bibitem{Czirok1997}
Czir\'ok A, Stanley H~E and Vicsek T 1997 {\em J. Phys. A\/} {\bf 30} 1375

\bibitem{Gregoire2004}
Gr\'egoire G and Chat\'e H 2004 {\em Phys. Rev. Lett.\/} {\bf 92} 025702

\bibitem{Chate2008}
Chat\'e H, Ginelli F, Gr\'egoire G and Raynaud F 2008 {\em Phys. Rev. E\/} {\bf
  77} 046113

\bibitem{Csahok2002}
Csah\`ok Z and Czir\`ok A 2002 {\em Physica A\/} {\bf 243} 304

\bibitem{Toner1995}
Toner J and Tu Y 1995 {\em Phys. Rev. Lett.\/} {\bf 75} 4326

\bibitem{Toner1998}
Toner J and Tu Y 1998 {\em Phys. Rev. E\/} {\bf 58} 4828--4858

\bibitem{Bertin2006}
Bertin E, Droz M and Gr\'egoire G 2006 {\em Phys. Rev. E\/} {\bf 74} 022101

\bibitem{Baskaran2008}
Baskaran A and Marchetti M 2008 {\em Phys. Rev. Lett.\/} {\bf 101} 268101

\bibitem{Dauxois}
Dauxois T and Peyrard M 2006 {\em Physics of solitons\/} (Cambridge: Cambridge
  University Press)

\bibitem{Albano}
Albano E~V 1996 {\em Phys. Rev. Lett.\/} {\bf 77} 2129

\bibitem{Aldana2007}
Aldana M, Dossetti V, Huepe C, Kenkre V~M and Larralde H 2007 {\em Phys. Rev.
  Lett.\/} {\bf 98} 095702

\bibitem{Huepe}
Aldana M and Huepe C 2003 {\em J. Stat. Phys.\/} {\bf 112} 135

\bibitem{Couzin2002}
Couzin I~D 2002 {\em J. Theor. Biol.\/} {\bf 218} 1

\bibitem{Duparcmeur}
Duparcmeur Y~L, Herrman H and Troadec J~P 1995 {\em J. Phys. I France\/} {\bf
  5} 1119

\bibitem{Gregoire2001}
Gr\'egoire G, Chat\'e H and Tu Y 2001 {\em Phys. Rev. E\/} {\bf 64} 011902

\bibitem{Levine}
Levine H, Rappel W~J and Cohen I 2000 {\em Phys. Rev. E\/} {\bf 63} 017101

\bibitem{Mikhailov}
Mikhailov A~S and Zanette D~H 1999 {\em Phys. Rev. E\/} {\bf 60} 4571

\bibitem{Ken}
Shimoyama N, Sugawara K, Mizuguchi T, Hayakawa Y and Sano M 1996 {\em Phys.
  Rev. Lett.\/} {\bf 76} 3870

\bibitem{Ratushnaya}
Ratushnaya V~I, Bedeaux D, Kulinskii V~L and Zvelindovsy A~V 2007 {\em Physica
  A\/} {\bf 381} 39--46

\bibitem{Degond}
Degond P and Motsch S 2007 {\em Comptes-rendus Math\'ematiques\/} {\bf 345}
  555--560

\bibitem{Degond2008}
Degond P and Motsch S 2008 {\em Journal of Statistical Physics\/} {\bf 131}
  989--1021

\bibitem{Binder}
Binder K 1976 {\em Phase transitions and critical phenomena\/} ed Domb C and
  Green M~S (Academic Press)

\bibitem{Muller}
M\"uller-Krumbhaar H and Binder K 1973 {\em J Stat. Phys.\/} {\bf 8} 1

\bibitem{Privman}
Privman V (ed) 1990 {\em Finite size scaling and numerical simulations of
  statistical systems\/} (Singapor: ed. World scientific)

\bibitem{Borgs}
Borgs C and Koteck\`y R 1990 {\em J Stat. Phys.\/} {\bf 61} 79

\bibitem{Lubeck}
L\"ubeck S 2004 {\em Int. J. of Mod. Phys. B\/} {\bf 18} 3977--4118

\bibitem{Marcq}
Marcq P, Chat\'e H and Manneville P 2006 {\em Prog. Theor. Phys. suppl\/} {\bf
  161} 244

\bibitem{These_gg}
Gr\'egoire G 2002 {\em Mouvement collectif et physique hors d'\'equilibre\/}
  Ph.D. thesis Universit\'e Paris 7--Denis Diderot

\bibitem{Baglietto}
Baglietto G and Albano E~V 2008 {\em Phys. Rev. E\/} {\bf 78} 021125

\end{thebibliography}
\bibliographystyle{iopart-num}

\end{document}